\documentclass[aps,pra,twocolumn,groupedaddress,longbibliography]{revtex4-2}
\usepackage{amsmath,amsfonts,amssymb}
\usepackage[english]{babel}
\usepackage{physics}
\usepackage{bm}
\usepackage{hyperref}
\usepackage{graphicx}
\usepackage{tikz}
\usepackage{placeins}
\usepackage{float}          
\usepackage{float}
\usepackage{subcaption}
\makeatletter
\let\newfloat\newfloat@ltx
\makeatother
\usepackage{algorithmicx}   
\usepackage{algcompatible}  

\floatstyle{ruled}
\newfloat{algorithm}{tbp}{loa}
\floatname{algorithm}{Algorithm}

\usepackage[title]{appendix}

\newcommand{\uv}{Departament de Física Teòrica and IFIC, Universitat de València-CSIC, 46100 Burjassot (València), Spain}

\begin{document}

\title{Pauli Propagation for Imaginary-Time Evolution}

\author{Rafael Gómez-Lurbe}
\email{rafael.gomez-lurbe@uv.es}
\author{Armando Pérez}
\affiliation{\uv}

\begin{abstract}
We extend the Pauli Propagation framework to imaginary-time evolution and introduce imaginary-time Pauli Propagation (ITPP), an operator-based algorithm for approximating thermal and ground-state properties directly in the Pauli basis. We derive explicit imaginary-time propagation rules for Pauli strings and analyze the main approximation errors arising from Trotterization and Pauli-space truncation, including bounds that account for the non-unitary nature of the evolution. Benchmarking ITPP on the one-dimensional transverse field Ising model, we find that the method performs best in the high-temperature regime, where thermal states admit a compressed Pauli-basis approximation, while the required number of Pauli terms grows rapidly as the system approaches the ground-state limit. We further compare with tensor network simulations and study finite-temperature scalability through the number of retained Pauli terms required to reach a fixed target accuracy. These results establish ITPP as a complementary framework for imaginary-time simulation and suggest that combining imaginary-time and real-time Pauli Propagation could provide a pathway toward simulating more general non-unitary and open quantum system dynamics within a unified Pauli-based framework.
\end{abstract}

\maketitle

\section{Introduction} \label{sec:introduction}

Classical simulation of quantum many-body systems is notoriously challenging due to the exponential growth of the Hilbert-space dimension with system size, rendering exact diagonalization rapidly intractable. As a consequence, approximate numerical methods are indispensable. Established techniques such as tensor network algorithms~\cite{Schollw_ck_2011,M_ller_Hermes_2012,Or_s_2014,Ba_uls_2023} and quantum Monte Carlo~\cite{RevModPhys.73.33,Assaad2008} have enabled substantial progress, yet they also face well-known limitations. In particular, tensor-network methods are generally expected to become more challenging in higher-dimensional or highly connected systems, where for fixed bond dimension the computational cost can grow exponentially with lattice connectivity. Quantum Monte Carlo methods, on the other hand, suffer from the sign problem in frustrated or fermionic models~\cite{PhysRevB.41.9301,PhysRevLett.94.170201}. These limitations motivate the exploration of alternative classical simulation frameworks.

A recently introduced approach for approximately simulating quantum dynamics is \emph{Pauli Propagation}~\cite{PhysRevA.99.062337,rudolph2023classicalsurrogatesimulationquantum}, also known as \emph{sparse Pauli dynamics}~\cite{Begu_i__2024}. In contrast to conventional Schrödinger-picture simulations that evolve a quantum state, Pauli Propagation works in the Heisenberg picture by expanding observables in the Pauli operator basis and evolving their coefficients in time. Operator growth is controlled by truncating the Pauli expansion and retaining only the most significant terms. This strategy is computationally attractive because Pauli operators admit efficient bitstring representations and their multiplication and commutation relations reduce to simple bitwise operations. Using this framework, state-of-the-art classical simulations of real-time quench dynamics in two- and three-dimensional transverse-field Ising models have been demonstrated~\cite{PRXQuantum.6.020302}. Furthermore, Ref.~\cite{lh6x-7rc3} showed that Pauli Propagation can also be used to classically estimate expectation values of arbitrary observables in random unstructured quantum circuits. More broadly, Pauli Propagation provides a complementary perspective to tensor-network methods: instead of controlling complexity via entanglement and low-rank structure, it exploits sparsity in operator space.

The original formulation of Pauli Propagation applies to real-time dynamics. In this work, we extend Pauli Propagation to imaginary-time evolution. \emph{Imaginary-time evolution} (ITE) is a fundamental tool for state preparation, enabling access to thermal states and filtering toward ground states, and is routinely implemented in classical simulation methods such as quantum Monte Carlo~\cite{RevModPhys.73.33} and tensor-network algorithms~\cite{Schollw_ck_2011}. Our approach generalizes Pauli Propagation to imaginary-time evolution and provides an operator-based framework for approximating thermal and ground-state observables directly in the Schr\"odinger picture. More broadly, the ability to handle non-unitary imaginary-time evolution within the Pauli Propagation framework suggests a possible route toward combining imaginary- and real-time dynamics, with potential applications to the simulation of open quantum system dynamics~\cite{PRXQuantum.3.010320}. An important advantage of approximating the density matrix and working directly in the Schr\"odinger picture is that a single propagated state can be reused to compute many different observables and thermodynamic quantities. This contrasts with the standard real-time Pauli Propagation framework, where one typically evolves observables in the Heisenberg picture and therefore targets the expectation value of a specific observable at a time.

Our approach is complementary to existing classical and quantum methods for imaginary-time evolution. Classical finite-temperature techniques such as quantum Monte Carlo and tensor-network imaginary-time evolution control complexity through stochastic sampling or low-entanglement representations, respectively, whereas the present method exploits sparsity in the Pauli expansion of the density matrix. It therefore provides an alternative operator-based route to accessing thermal and ground-state properties, with a distinct set of trade-offs. In particular, ITPP may serve as a useful baseline for further algorithmic developments and could be relevant in regimes where conventional methods face challenges, such as higher-dimensional systems that are difficult for tensor-network methods or models affected by sign-problem limitations in Monte Carlo simulations. Quantum imaginary-time algorithms, such as quantum imaginary-time evolution (QITE)~\cite{Motta_2019} and its variational variant~\cite{McArdle_2019}, aim to approximate imaginary-time dynamics directly on quantum hardware. While these approaches are promising, their near-term implementation remains challenging due to measurement overheads, finite coherence times, and hardware noise. As a result, classical methods remain essential both as practical simulation tools and as benchmarks for quantum algorithms. From this perspective, ITPP provides a classical point of comparison in regimes where truncated Pauli representations remain effective.

As in Pauli Propagation for real-time dynamics, our imaginary-time formulation relies on a Trotter decomposition of the evolution operator, a standard ingredient in many imaginary-time simulation schemes. The overall accuracy of the method is therefore governed by two distinct approximations: the Trotterization of the imaginary-time evolution and the truncation of the Pauli expansion required to control operator growth. In this work, we discuss both sources of error and their implications for the practical performance of the algorithm.

The basic intuition behind our approach is that finite-temperature states can be much more compact in the Pauli basis than zero-temperature pure states. At one extreme, the infinite-temperature state is the maximally mixed state, which is represented by a single Pauli operator, namely the identity. At the other extreme, a generic ground state is pure and typically exponentially dense in the Pauli basis, generally requiring at least \(2^N\) Pauli operators for an \(N\)-spin system. This suggests that, starting from the maximally mixed state and evolving in imaginary time, there should exist a finite-temperature regime in which the density matrix can still be accurately approximated by a sparse Pauli expansion. From this perspective, Pauli Propagation is naturally expected to be most effective when the density matrix admits an accurate sparse approximation in the Pauli basis, meaning that most of the weight of its Pauli coefficients is concentrated on a relatively small subset of Pauli strings, as is expected at sufficiently high temperatures. Conversely, as the temperature is lowered and the system approaches the ground state, operator growth is expected to become more pronounced, so that the number of Pauli terms required to maintain a given accuracy may increase rapidly, thereby limiting the efficiency of the method.

In this work, we derive explicit update rules for Pauli propagation under imaginary-time evolution and develop the imaginary-time Pauli Propagation (ITPP) algorithm for approximating thermal and ground-state properties. We analyze the approximation errors introduced by Trotterization and Pauli-space truncation, and investigate the practical performance of the method through numerical simulations.

As a benchmark, we apply ITPP to the one-dimensional transverse-field Ising model in the ordered phase. We compare the results with Trotterized imaginary-time evolution, exact calculations, and established classical tensor-network techniques for finite-temperature systems, specifically matrix-product density operators (MPDOs)~\cite{Verstraete_2004,de_las_Cuevas_2020} evolved using the time-evolving block decimation (TEBD)~\cite{Vidal_2003} method. We also examine several truncation strategies, including coefficient-threshold and fixed-\(K\) schemes. Our results show that ITPP accurately approximates high-temperature thermal-state properties with relatively modest resources, while substantially more resources are required as the temperature decreases, in agreement with the sparsity intuition underlying the method.

The paper is organized as follows. In Sec.~\ref{sec:theory}, we briefly review imaginary-time evolution, thermal states, and the standard Pauli Propagation algorithm for real-time dynamics. Next, we introduce our extension of Pauli Propagation to imaginary-time dynamics. Numerical experiments are presented in Sec.~\ref{sec:experiments}. Finally, Sec.~\ref{sec:conclusion} summarizes our findings and discusses possible future directions.

\section{Theory}\label{sec:theory}

In this section, we review the theoretical background underlying our approach. We begin by summarizing imaginary-time evolution and its connection to thermal and ground states. We then outline the standard Pauli Propagation algorithm for real-time dynamics. Finally, we introduce our extension of Pauli Propagation to imaginary-time evolution, which forms the basis of the method used in this work.

\subsection{Imaginary-Time Evolution}

Imaginary-time evolution is a useful framework for extracting ground-state properties of quantum many-body systems. For a Hamiltonian \(H\) and an initial pure state \(\ket{\psi_0}\), it is defined by the non-unitary transformation
\begin{equation}
    \ket{\psi(\tau)} =
    \frac{e^{-\tau H}\ket{\psi_0}}
    {\sqrt{\bra{\psi_0} e^{-2\tau H} \ket{\psi_0}}},
    \label{eq:ITE_state}
\end{equation}
where \(\tau \ge 0\) denotes imaginary time. The normalization in Eq.~\eqref{eq:ITE_state} ensures that \(\ket{\psi(\tau)}\) remains a valid quantum state.

Expanding \(\ket{\psi_0}\) in the energy eigenbasis of \(H\), each component is weighted by a factor \(e^{-\tau E_n}\), so contributions from higher-energy states are exponentially suppressed as \(\tau\) increases. In the non-degenerate case, if \(\ket{\psi_0}\) has nonzero overlap with the ground state, the evolved state converges to the unique ground state of \(H\). If the ground-state subspace is degenerate, the long-imaginary-time limit instead becomes the normalized projection of \(\ket{\psi_0}\) onto that subspace, determined by the initial overlaps. For simplicity, we focus in the following on the non-degenerate case.

More generally, imaginary-time evolution can be formulated at the level of density matrices.
Given an initial density matrix $\rho_0$, its imaginary-time evolved state is defined as
\begin{equation}
    \rho(\tau) =
    \frac{e^{-\tau H} \, \rho_0 \, e^{-\tau H}}
    {\operatorname{Tr}\!\left[e^{-2\tau H} \rho_0\right]}.
    \label{eq:ITE_density}
\end{equation}
This expression ensures proper normalization, $\operatorname{Tr}[\rho(\tau)] = 1$, for all imaginary times.

In the limit $\tau \to \infty$, and provided that $\rho_0$ has a nonzero overlap with the ground-state subspace, the density matrix $\rho(\tau)$ converges to a projector onto that subspace. In the non-degenerate case, this reduces to
\begin{equation}
    \rho(\tau) \xrightarrow{\tau \to \infty}
    \ket{\psi_{\mathrm{gs}}}\!\bra{\psi_{\mathrm{gs}}},
\end{equation}
where $\ket{\psi_{\mathrm{gs}}}$ denotes the ground state of $H$.

A particularly important case is obtained by choosing the initial state to be the maximally mixed state,
\[
\rho_0 = \frac{1}{2^n}\,\mathbb{I},
\]
where \(n\) denotes the number of spins in the system.

In this case, the imaginary-time evolution defined in Eq.~\eqref{eq:ITE_density} yields, for an arbitrary imaginary-time \(\tau\), the thermal state
\begin{equation}
    \rho_{\mathrm{th}}(\tau) =
    \frac{e^{-2\tau H}}
    {\operatorname{Tr}\!\left[e^{-2\tau H}\right]},
    \label{eq:Thermal_State}
\end{equation}
obtained by evolving the infinite-temperature state at \(\tau = 0\) to imaginary-time \(\tau\), which sets the inverse temperature \(\beta=2\tau\). As discussed above, in the zero-temperature limit, \(\tau \to \infty\), the imaginary-time evolution converges to a projector onto the ground state.

Expectation values of observables $O$ at imaginary-time \(\tau\) are obtained as
\begin{equation}
    \langle O \rangle_\tau = \operatorname{Tr}\!\left[O \, \rho_{\mathrm{th}}(\tau)\right]
    = \frac{\operatorname{Tr}\!\left[O \, e^{-2\tau H}\right]}
    {\operatorname{Tr}\!\left[e^{-2\tau H}\right]}.
    \label{eq:Expectation_Value}
\end{equation}
In the zero-temperature limit $\tau \to \infty$, the thermal state $\rho_{\mathrm{th}}(\tau)$ projects onto the ground state of $H$, and the expectation value $\langle O \rangle_\tau$ reduces to the ground-state expectation value of $O$.

Several classical approaches have been developed to approximate imaginary-time evolution, most notably quantum Monte Carlo~\cite{RevModPhys.73.33} and tensor-network algorithms~\cite{Schollw_ck_2011}. In recent years, quantum algorithms for approximating imaginary-time evolution on quantum hardware have also been proposed. Approaches such as Quantum imaginary-time Evolution (QITE)~\cite{Motta_2019} and its variational variant (VQITE)~\cite{McArdle_2019} implement imaginary-time evolution directly on a quantum computer in the Schrödinger picture. For a pedagogical introduction to these methods, see Ref.~\cite{angléscastillo2025understandingquantumimaginarytime}.

\subsection{Pauli Propagation}

Pauli Propagation is a classical algorithm for simulating quantum dynamics by evolving observables in the Heisenberg picture, rather than state vectors. In this framework, an observable $O$ evolves as
\begin{equation}
    O(t) = U^\dagger(t) \, O \, U(t),
\end{equation}
with $U(t) = \exp(-i t H)$, where $H$ is the system Hamiltonian.

The key idea is to expand the observable in the Pauli basis:
\begin{equation}
    O = \sum_{P \in \mathcal{P}} c_P P,
    \label{eq:Observable_Pauli}
\end{equation}
where $\mathcal{P}$ is a set of Pauli operators and $c_P$ are real coefficients. The evolution of a Pauli operator \(P\) under a Pauli rotation $U_Q(\theta) = \exp(-i Q \theta / 2)=\cos\!\left(\frac{\theta}{2}\right)\mathbb{I}
- i \sin\!\left(\frac{\theta}{2}\right) Q$ satisfies
\begin{equation}
    \begin{aligned}
        U_Q(\theta)^\dagger & P \, U_Q(\theta) \\
        = {} &
        \begin{cases}
            P, & [P, Q] = 0, \\[4pt]
            \cos(\theta)\, P + i \sin(\theta)\, Q P, & \{P, Q\} = 0 ,
        \end{cases}
    \end{aligned}
\end{equation}
where $[\,\cdot\,,\,\cdot\,]$ and $\{\,\cdot\,,\,\cdot\,\}$ denote the commutator and anticommutator, respectively. This identity enables each term in the Pauli decomposition of Eq.~\ref{eq:Observable_Pauli} to be propagated independently.

For a Hamiltonian decomposed as
\begin{equation}
    H = \sum_{j=1}^{L} \alpha_j h_j,
\end{equation}
where the \(h_j\) are Pauli operators, the exact real-time evolution operator is
\begin{equation}
    U(t) = e^{-iHt}.
\end{equation}
To approximate \(U(t)\), we employ a first-order Trotter expansion. Writing
\(t = n\,\Delta t\), the Trotterized evolution operator is defined as
\begin{equation}
    U_{\mathrm{Trotter}}(t)
    =
    \left(
        \prod_j \exp(-i \alpha_j \Delta t\, h_j)
    \right)^n,
\end{equation}
where \(n=t/\Delta t\). This yields a first-order approximation to \(U(t)\) in the sense that
\begin{equation}
    U(t)=U_{\mathrm{Trotter}}(t)+\mathcal{O}\!\left(C\,t\,\Delta t\right),
\end{equation}
where the error term is understood in operator norm.

The local error per Trotter step is of order \(\mathcal{O}(\Delta t^2)\). Over \(n\) steps, this accumulates into a global error of order
\begin{equation}
    \mathcal{O}\!\left(C\,t\,\Delta t\right).
\end{equation}
Here,
\begin{equation}
    C := \sum_{j<k} |\alpha_j\alpha_k|\,\|[h_j,h_k]\|
\end{equation}
quantifies the non-commutativity of the Hamiltonian terms. In many practical settings, the total evolution time \(t\) is fixed while the step size \(\Delta t\) is varied. In this regime, the global error therefore scales linearly with \(\Delta t\). This decomposition expresses \(U(t)\) as a sequence of single-Pauli rotations, which can then be applied iteratively to propagate the observable.

A challenge of Pauli Propagation is the exponential growth of terms in $O(t)$, as each rotation can generate new Pauli strings. To maintain the computational cost tractable, two main truncation strategies are employed:
\begin{itemize}
    \item \textbf{Coefficient threshold:} retain only Pauli terms with coefficients satisfying $|c_P| > \delta$, where $\delta$ is a chosen threshold parameter.
    \item \textbf{Pauli weight cutoff:} retain terms whose Pauli weight (number of non-identity operators) does not exceed a fixed threshold.
\end{itemize}

Coefficient-based truncation is practical and widely used~\cite{PRXQuantum.6.020302}, while weight-based truncation allows rigorous bounds on computational cost and approximation error~\cite{lh6x-7rc3}. Each method balances efficiency and accuracy differently, depending on the system and observable structure.

\subsection{Imaginary-Time Evolution via Pauli Propagation}\label{Sec:ITPP}

Following a similar approach to real-time Pauli Propagation, we define a set of rules to evolve a Pauli operator under an imaginary-time Pauli operator
\begin{equation}
    V_Q(\tau) = \exp\left(-\frac{\tau}{2} Q\right)=\cosh\!\left(\frac{\tau}{2}\right)\mathbb{I}
-  \sinh\!\left(\frac{\tau}{2}\right) Q,
\end{equation}
where $Q$ is a Pauli string.  

The evolution of a Pauli operator $P$ under this imaginary-time operator satisfies

\begin{equation}
    \begin{aligned}
        V_Q(\tau)^\dagger & P \, V_Q(\tau) \\
        = {} &
        \begin{cases}
        P, & \{P, Q\} = 0, \\[4pt]
        \cosh(\tau)\, P - \sinh(\tau)\, Q P, & [P, Q] = 0 .
        \end{cases}
    \end{aligned}
    \label{eq:ITPP_PropRules}
\end{equation}

The derivation of these propagation rules is provided in Appendix~\ref{app:ITPP_propagation_rules}.

Interestingly, the propagation rules for a Pauli string under imaginary-time evolution are essentially the opposite of those encountered in real-time Pauli Propagation. In particular, when the propagated Pauli string anticommutes with the generator of the imaginary-time operator, the operator remains unchanged. By contrast, when the two operators commute, the Pauli string splits into two terms.

This behavior contrasts with the real-time case, where nontrivial evolution arises from anticommuting generators and the resulting coefficients are trigonometric functions reflecting unitarity and norm preservation. In imaginary-time evolution, the coefficients are instead hyperbolic functions, corresponding to the non-unitary nature of the dynamics.

Together, these rules provide a natural extension of Pauli Propagation to imaginary-time evolution, enabling the estimation of thermal and ground-state properties within the Pauli operator framework.

Similar to real-time evolution, we consider a Hamiltonian decomposed as
\begin{equation}
    H = \sum_{j=1}^{L} \alpha_j h_j,
\end{equation}
where \(h_j\) are Pauli operators. The exact imaginary-time evolution operator is
\begin{equation}
    V(\tau) = e^{-\frac{\tau}{2} H}.
\end{equation}
We include the factor of \(1/2\) in the exponent so that, when applied to
the maximally mixed initial state as
\(V(\tau)\,\frac{\mathbb{I}}{2^n}\,V^\dagger(\tau)\), the parameter \(\tau\)
coincides with the inverse temperature \(\beta\) of the resulting thermal state.

To approximate \(V(\tau)\), we use a first-order Trotter expansion. Higher-order product formulas, such as the second-order Strang splitting, could also be used to reduce the Trotter error. For example, replacing a first-order decomposition by a symmetric product formula of the form
\(e^{-\frac{\Delta\tau}{2}A} e^{-\Delta\tau B} e^{-\frac{\Delta\tau}{2}A}\)
increases the order of the approximation with only a modest increase in computational cost
\cite{Hadfield_2018,PhysRevX.11.011020}. Here we focus on the first-order formula as a baseline, in order to isolate the effects of Pauli propagation and truncation. The extension to higher-order product formulas is straightforward within the present framework and is left for future work. 
We therefore write \(\tau = n\,\Delta\tau\) and define the Trotterized evolution operator as
\begin{equation}
    V_{\mathrm{Trotter}}(\tau)
    =
    \left(
        \prod_j \exp\!\left(- \frac{\alpha_j}{2} \Delta\tau\, h_j\right)
    \right)^n ,
\end{equation}
where \(n=\tau/\Delta\tau\). This yields a first-order approximation to \(V(\tau)\) in the sense that
\begin{equation}
    V(\tau)=V_{\mathrm{Trotter}}(\tau)
    + \mathcal{O}\!\left(e^{\frac{\tau}{2}\mu}\, C\,\tau\,\Delta\tau\right),
\end{equation}
where the error term is understood in operator norm.

The local error per Trotter step is of order \(\mathcal{O}(\Delta\tau^2)\). Over \(n\) steps, this accumulates into a global error of order
\begin{equation}
    \mathcal{O}\!\left(e^{\frac{\tau}{2}\mu}\, C\,\tau\,\Delta\tau\right).
\end{equation}
In contrast to the unitary real-time case, this bound contains the exponential factor \(e^{\tau\mu/2}\), which reflects the non-unitary nature of imaginary-time evolution. Here,
\begin{equation}
    C := \sum_{j<k} |\alpha_j \alpha_k|\,\|[h_j,h_k]\|
\end{equation}
quantifies the non-commutativity of the Hamiltonian terms, while
\begin{equation}
    \mu := \sum_{j=1}^{L} |\alpha_j|
\end{equation}
captures the dependence on the Hamiltonian coefficients. A derivation of this bound is given in Appendix~\ref{app:trotter_imag_time}.

When the total imaginary time \(\tau\) is held fixed, the step size \(\Delta\tau\) becomes the parameter controlling the quality of the approximation. The global error then scales linearly with \(\Delta\tau\), so that
\begin{equation}
    V(\tau)=V_{\mathrm{Trotter}}(\tau)+\mathcal{O}(\Delta\tau),
\end{equation}
again with the error understood in operator norm.

To construct thermal states, we use the imaginary-time evolution operator. 
Given the exact operator \(V(\tau)=e^{-\frac{\tau}{2}H}\), the corresponding thermal state at inverse temperature \(\tau\) can be written as
\begin{equation}
    \rho_{\mathrm{th}}\!\left(\tfrac{\tau}{2}\right)
    =
    \frac{
        V^{\dagger}(\tau)\,\rho_0\,V(\tau)
    }{
        \Tr\!\left[V^{\dagger}(\tau)\,\rho_0\,V(\tau)\right]
    },
\end{equation}
where \(\rho_0 = \frac{\mathbb{I}}{2^n}\).

Using the first-order Trotter approximation \(V_{\mathrm{Trotter}}(\tau)\), we define the corresponding Trotterized thermal state
\begin{equation}
    \rho_{\mathrm{Trotter}}\!\left(\tfrac{\tau}{2}\right)
    :=
    \frac{
        V_{\mathrm{Trotter}}^{\dagger}(\tau)\,\rho_0\,V_{\mathrm{Trotter}}(\tau)
    }{
        \Tr\!\left[
            V_{\mathrm{Trotter}}^{\dagger}(\tau)\,\rho_0\,V_{\mathrm{Trotter}}(\tau)
        \right]
    }.
    \label{eq:ThermalTrotter}
\end{equation}

From the analysis in Appendix~\ref{app:trotter_imag_time_norm}, the error between the exact and Trotterized thermal states satisfies
\begin{equation}
    \|\rho_{\mathrm{th}} - \rho_{\mathrm{Trotter}}\|_1
    =
    \mathcal{O}\!\left(
        e^{\tau \lambda_{\max}(H)}
        e^{\frac{\tau}{2}\mu}\,
        C\,\tau\,\Delta\tau
    \right),
\end{equation}
where
\begin{equation}
    C := \sum_{j<k} |\alpha_j \alpha_k|\,\|[h_j,h_k]\|,
    \qquad
    \mu := \sum_{j=1}^{L} |\alpha_j|.
\end{equation}
In particular, when the total imaginary time \(\tau\) is fixed, this reduces to
\begin{equation}
    \|\rho_{\mathrm{th}} - \rho_{\mathrm{Trotter}}\|_1
    =
    \mathcal{O}(\Delta\tau).
\end{equation}

In practice, we approximate \(\rho_{\mathrm{Trotter}}\) as defined in Eq.~\eqref{eq:ThermalTrotter} by propagating the maximally mixed state
\(\rho_0 = \frac{\mathbb{I}}{2^n}\) under the Trotterized imaginary-time evolution operator.
Because the factor \(\frac{1}{2^n}\) appears in both the numerator and denominator, it cancels upon normalization. Consequently, one may equivalently propagate the identity operator \(\mathbb{I}\) directly, obtaining the same normalized state.

This propagation can be carried out using the rules introduced in
Eq.~\eqref{eq:ITPP_PropRules}. In the first step, the identity operator is
propagated through the initial imaginary-time operator. Since the identity
commutes with any Pauli string, this operation preserves the identity while
generating an additional Pauli string associated with the generator of the
imaginary-time evolution. For generality, we therefore express the resulting
operator as
\[
\rho = \sum_\alpha c_\alpha P_\alpha,
\]
a linear combination of Pauli strings \(P_\alpha\).

We then propagate \(\rho\) under the next imaginary-time step generated by a
Pauli string \(Q\),
\begin{equation}
    V_Q(\tau) = \exp\!\left(-\frac{\tau}{2} Q\right),
\end{equation}
leading to the transformation
\begin{equation}
    \rho \;\longrightarrow\; V_Q^\dagger(\tau)\, \rho\, V_Q(\tau).
\end{equation}

Since \(\rho\) is a sum of Pauli strings, the propagation rules
(Eq.~\eqref{eq:ITPP_PropRules}) can be applied independently to each term
\(P_\alpha\). This produces an updated operator that is again a linear
combination of Pauli strings. The operator is then re-expanded in the Pauli
basis and a truncation scheme is applied to limit the growth of the number of
Pauli terms, thereby controlling computational cost. In addition, the operator is normalized by its trace after each elementary propagation step. This normalization is computationally inexpensive, since in the Pauli representation it only requires identifying the coefficient of the identity operator and rescaling all stored coefficients accordingly. We find this per-step normalization to be important for numerical stability. In particular, when normalization is postponed until the end of a full Trotter layer, intermediate coefficients can grow significantly in larger spin systems, occasionally leading to numerical overflow. This behavior is consistent with the accumulation of contributions to the identity component during non-unitary imaginary-time propagation. Normalizing after each elementary step keeps the coefficient scale controlled and improves the stability of the subsequent truncation procedure. This normalization is not an additional approximation, but a rescaling of the operator representation.

The resulting operator serves as the input for the next imaginary-time
operation in the Trotter decomposition. Repeating this procedure for all
Pauli terms in each Trotter step, and iterating over the full sequence of
Trotter steps, defines the ITPP algorithm and yields an approximate
implementation of imaginary-time evolution within the Pauli propagation
framework. A step-by-step description of ITPP is provided in
Algorithm~\ref{algITPP}.

\begin{algorithm}[t]
    \caption{Imaginary-Time Pauli Propagation (ITPP)}
    \label{algITPP}
    \begin{algorithmic}[1]
    
    \STATE \textbf{Input:} Hamiltonian \(H = \sum_k h_k Q_k\); step size \(\Delta\tau\); final imaginary-time \(\tau\)
    
    \STATE Number of Trotter steps: \(N_\tau = \left\lceil \tau / \Delta\tau \right\rceil\)
    
    \STATE Trotterize imaginary-time evolution:
    \[
    e^{-\frac{\tau}{2} H} \approx \left( \prod_k e^{-\frac{\Delta\tau}{2} h_k Q_k} \right)^{N_\tau}
    \]
    
    \STATE Initialize \(\rho =\mathbb{I}\).
    
    \FOR{Trotter step \(n = 1, \dots, N_\tau\)}
        \FOR{each Pauli term \(h_k Q_k\) in the chosen Trotter ordering}
            \STATE \(\rho' \leftarrow 0\)
            \FOR{each Pauli term \(c_\alpha P_\alpha\) in \(\rho\)}
                \STATE Propagate using ITPP rules:
                \[
                P_\alpha \rightarrow V_{Q_k}^\dagger P_\alpha V_{Q_k}
                \]
                \STATE Accumulate resulting Pauli strings into \(\rho'\)
            \ENDFOR
            \STATE Truncate and normalize \(\rho \leftarrow \mathcal{T}(\rho')/\mathrm{Tr}(\mathcal{T}(\rho'))\)
        \ENDFOR
    \ENDFOR
    
    \STATE \textbf{Output:} propagated operator \(\rho\)
    
    \end{algorithmic}
\end{algorithm}

In ITPP, the overall approximation error has two main contributions: the Trotter error, which we have already bounded, and the truncation error introduced by restricting the Pauli expansion. In general, controlling the truncation error rigorously is challenging for structured Hamiltonians, especially when truncation is performed by thresholding coefficients, which is typically preferred in practical implementations due to its empirical performance.

Recent work has made progress in establishing rigorous bounds for truncation based on Pauli weight. In particular, Ref.~\cite{rudolph2026thermalstatesimulationpauli} derives explicit error bounds for first-order Trotterized imaginary-time evolution under weight truncation (see Theorem~3 therein). Under assumptions of bounded locality and degree, they show that there exist constants \(c_1,c_2>0\) such that the error in expectation values satisfies
\[
\bigl|\Tr\!\bigl[O(\rho-\tilde\rho)\bigr]\bigr|
\;\le\;
\|O\|_1\;
\exp\!\bigl(c_1\,\beta M\bigr)\;
\Bigl(c_2\,\beta\,\ell\,w\Bigr)^{ k/w },
\]
where \(k\) is the maximum retained Pauli weight and \(\|O\|_1\) denotes the \(\ell_1\) norm of the Pauli coefficients of the observable. These results provide rigorous guarantees on how the truncation error depends on locality and the retained Pauli support.

The same work also establishes related bounds for a qDRIFT-based imaginary-time evolution scheme, in which Hamiltonian terms are sampled randomly rather than applied through a deterministic Trotter expansion. In this setting, the analysis can also be extended to small-angle truncation, which is closer in spirit to coefficient-based truncation. The authors note that qDRIFT primarily serves as an analytical tool for studying operator spreading and deriving truncation guarantees.

However, the truncation strategy considered in Ref.~\cite{rudolph2026thermalstatesimulationpauli} differs from the coefficient-based truncation employed in this work. Extending such rigorous guarantees to more general truncation schemes, and in particular to coefficient-threshold truncation, remains an open problem. A similar situation arises in real-time Pauli Propagation, where rigorous bounds are known only in restricted settings, such as random circuits with anticoncentration properties combined with weight-based truncation \cite{lh6x-7rc3}. More broadly, as discussed in Ref.~\cite{PRXQuantum.6.020302}, Pauli propagation applied to structured Hamiltonians with coefficient-based truncation should be regarded as a heuristic approach in the absence of general error guarantees.

Once the approximate state $\tilde{\rho}_{th}(\tau)$ is obtained, expectation
values of observables $O$ can be computed directly as
\begin{equation}
    \langle O \rangle_\tau \approx
    \operatorname{Tr}\!\left[O\,\tilde{\rho}_{th}(\tau)\right].
\end{equation}
This procedure enables the prediction of thermal or ground-state properties, while retaining the flexibility to evaluate multiple observables using the
same propagated operator. By adjusting the truncation threshold and the Trotter step size,
one can systematically tune the tradeoff between computational cost and accuracy, making this approach well suited for controlled and flexible studies of quantum many-body systems.

It is important to note that truncation in the Pauli basis does not, in general, preserve positive semidefiniteness. As a result, the approximate operator $\tilde{\rho}_{\mathrm{th}}(\tau)$ may develop negative eigenvalues and therefore fail to represent a valid quantum state. This can, in principle, lead to unphysical expectation values.

In practice, the loss of positive semidefiniteness arises from the truncation procedure, which modifies the operator spectrum in a non-controlled manner. For moderate truncation, these violations are typically small, for example resulting in slightly negative eigenvalues, and expectation values of local observables often remain accurate, as they are primarily determined by the dominant Pauli components. However, for more aggressive truncation, the deviation from a valid quantum state can become significant, and caution is required when interpreting results, especially for global quantities or when the approximate state is used in subsequent computations.

A straightforward way to restore positive semidefiniteness is to square the approximate density operator. This guarantees a positive semidefinite operator, but comes at a significant computational cost, as squaring substantially increases the Pauli support. The resulting operator must then be normalized by its trace to recover a properly normalized quantum state. Since this procedure effectively applies imaginary-time evolution twice, the squared operator corresponds to an approximate thermal state at inverse temperature $\beta = 2\tau$.

The computational overhead associated with explicitly enforcing positive semidefiniteness can be mitigated using the following identity. Here, \(\langle O \rangle_{2\tau}\) denotes the expectation value of the observable \(O\) at imaginary time \(2\tau\), and \(\tilde{\rho}_{\mathrm{th}}(\tau)\) is the ITPP approximation to the thermal state at imaginary time \(\tau\):
\begin{equation}
    \langle O \rangle_{2\tau} \approx
    \frac{\operatorname{Tr}\!\left[O\,\tilde{\rho}_{\mathrm{th}}^2(\tau)\right]}
         {\operatorname{Tr}\!\left[\tilde{\rho}_{\mathrm{th}}^2(\tau)\right]}
    =
    \frac{\operatorname{Tr}\!\left[\big(O\,\tilde{\rho}_{\mathrm{th}}(\tau)\big)
    \tilde{\rho}_{\mathrm{th}}(\tau)\right]}
         {\operatorname{Tr}\!\left[\tilde{\rho}_{\mathrm{th}}^2(\tau)\right]}.
\end{equation}

Rather than explicitly constructing $\tilde{\rho}_{\mathrm{th}}^2(\tau)$, one can instead compute the Pauli expansion of the product $O\,\tilde{\rho}_{\mathrm{th}}(\tau)$, which typically contains significantly fewer terms, and subsequently evaluate its overlap with $\tilde{\rho}_{\mathrm{th}}(\tau)$. The denominator corresponds to the purity of the approximate state and can be computed efficiently from the $\ell_2$ norm of the coefficient vector in the Pauli basis. This approach enables the evaluation of physically meaningful observables while avoiding the explicit construction of a squared operator with large Pauli support.

\paragraph{Practical considerations and truncation strategies.}
In practice, the efficiency of ITPP is largely determined by the truncation scheme used to control the growth of the Pauli expansion. In our numerical experiments, as in other Pauli Propagation studies, we find coefficient-threshold truncation to be the most effective strategy: Pauli terms whose coefficients fall below a prescribed threshold are discarded. Following common practice~\cite{PRXQuantum.3.010320}, we choose thresholds as negative powers of two and vary them systematically to assess the trade-off between accuracy and computational cost.

A limitation of coefficient-based truncation is that it does not provide direct control over the number of retained Pauli terms. As a result, the computational cost can vary significantly during the evolution, and in some cases the number of terms may grow rapidly. In practice, this can be mitigated by imposing an upper bound on the number of retained terms and terminating or adapting the simulation once this bound is reached.

To facilitate a more controlled analysis of scalability, it is also useful to consider an alternative truncation scheme based on a fixed number of retained Pauli terms. In this approach, which we refer to as $K$-truncation, only the $K$ largest coefficients (in magnitude) are kept after each propagation step. While this method is less flexible than threshold-based truncation, it provides direct control over the computational cost and is therefore convenient for systematic studies of scaling behavior. We emphasize, however, that $K$-truncation is primarily used here as a diagnostic tool, whereas coefficient-based truncation is more suitable for practical simulations.

\paragraph{Scope, limitations, and regimes of applicability.}
The performance of ITPP is ultimately governed by the growth of Pauli support during imaginary-time evolution. For local Hamiltonians at short to moderate imaginary times, corresponding to relatively high-temperature thermal states, the density matrix can often be well approximated by a sparse expansion in the Pauli basis. In this regime, accurate expectation values may be obtained with a modest number of retained terms. As the imaginary time increases and the system approaches lower temperatures, however, the Pauli expansion typically becomes increasingly dense, and the number of terms required to maintain a fixed accuracy can grow rapidly. In the limiting case \(\beta \to \infty\), the thermal state approaches the ground state. For a generic \(N\)-spin system, this is a pure state whose Pauli expansion can require at least \(2^N\) nonzero terms. This growth, which is also reflected in our numerical results, represents a fundamental limitation of the method.

More generally, the truncation error introduced by restricting the Pauli expansion is difficult to control rigorously for practical truncation schemes, particularly those based on coefficient thresholding. As discussed above, this limits the possibility of obtaining general a priori error guarantees, and the method should therefore be regarded as heuristic in such settings. From a computational-complexity perspective, this limitation is expected: preparing the ground state of a general local Hamiltonian is QMA-hard in the worst case~\cite{kitaev2002classical}, so no efficient classical method is expected to succeed uniformly at large imaginary times or very low temperatures. Thus, the breakdown of ITPP in this regime reflects the intrinsic difficulty of the underlying problem rather than a limitation specific to the implementation.


\begin{figure*}[t]
    \centering
    \textbf{Comparison of ITPP, Trotterized ITE, and Exact ITE} \vspace{0.2cm} \\
    \includegraphics[scale=0.32]{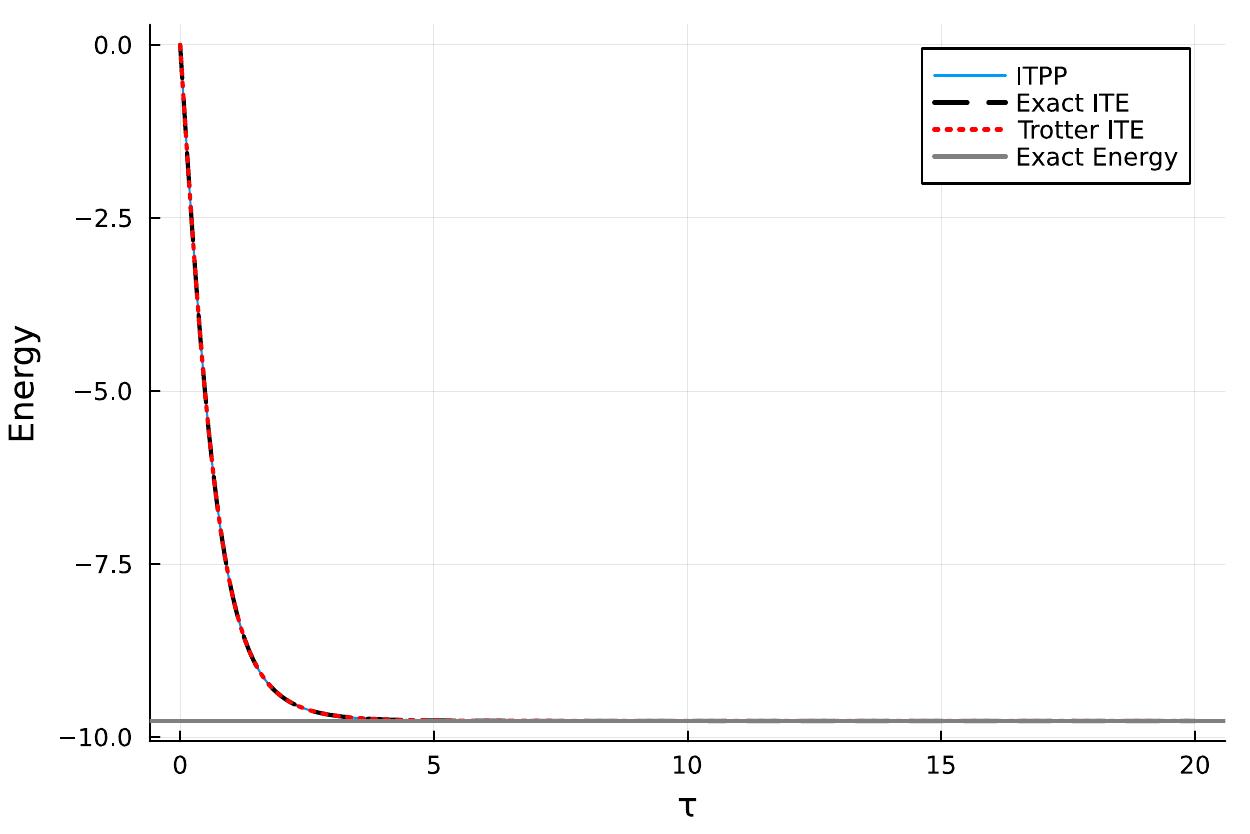}
    \includegraphics[scale=0.32]{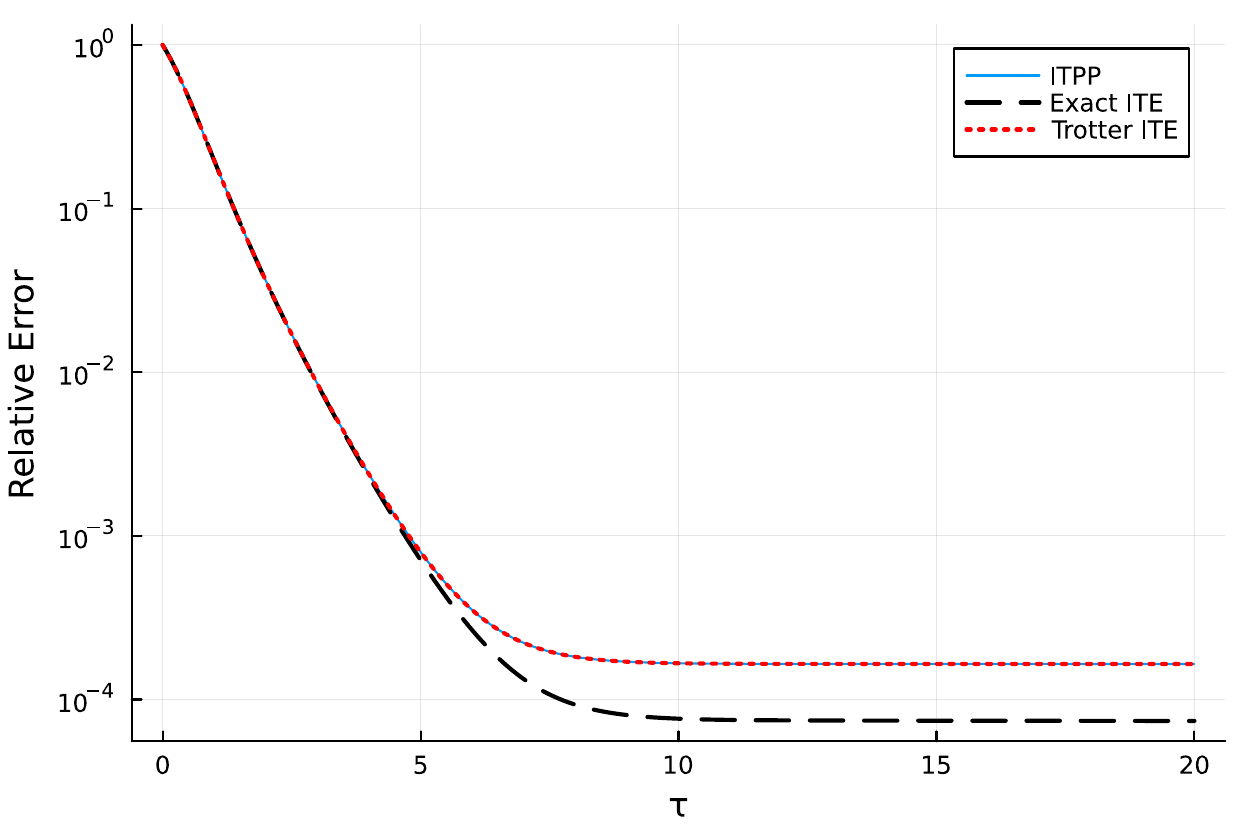}
    \caption{Comparison of ITPP with Trotterized and exact imaginary-time
        evolution for the one-dimensional transverse-field Ising model (TFIM) with
        \(N = 10\) spins, \(J = 1\), and \(h = 0.5\). The simulations use a Trotter
        step size \(\Delta \tau = 0.04\) and truncation threshold \(\delta = 0\).
        The left panel shows the energy as a function of imaginary time, while the
        right panel displays the corresponding relative error with respect to the
        exact ground-state energy computed using the Bogoliubov--de\,Gennes solution.
        In the left panel, the solid line indicates this exact ground-state energy.
        The relative error is shown on a logarithmic scale. The results demonstrate
        that ITPP faithfully reproduces Trotterized imaginary-time evolution.}

    \label{fig:ITE}
\end{figure*}


Within these limitations, ITPP provides a complementary approach to existing classical and quantum methods. It is best suited to regimes where the density matrix remains approximately sparse in the Pauli basis and where approximate expectation values of observables are sufficient. In such cases, a single propagated operator can be reused to evaluate multiple observables without rerunning the simulation. Unlike tensor-network methods, ITPP does not rely on low-entanglement structure, and unlike quantum Monte Carlo it is not formulated as a stochastic sampling method, so it offers a distinct set of trade-offs that may be useful in settings where those approaches face difficulties, such as highly connected or higher-dimensional systems, or models affected by sign problems. Compared with quantum imaginary-time algorithms, ITPP provides a fully classical benchmark and avoids the measurement overhead, noise, and finite-coherence constraints associated with near-term quantum hardware. The method is therefore not intended to replace existing techniques, but rather to provide an additional Pauli-basis framework for studying imaginary-time evolution.


\begin{figure*}
    \centering
    \textbf{ITPP performance for \(N=12\)} \vspace{0.2cm}

    \begin{subfigure}[t]{0.48\textwidth}
        \centering
        \includegraphics[width=\linewidth]{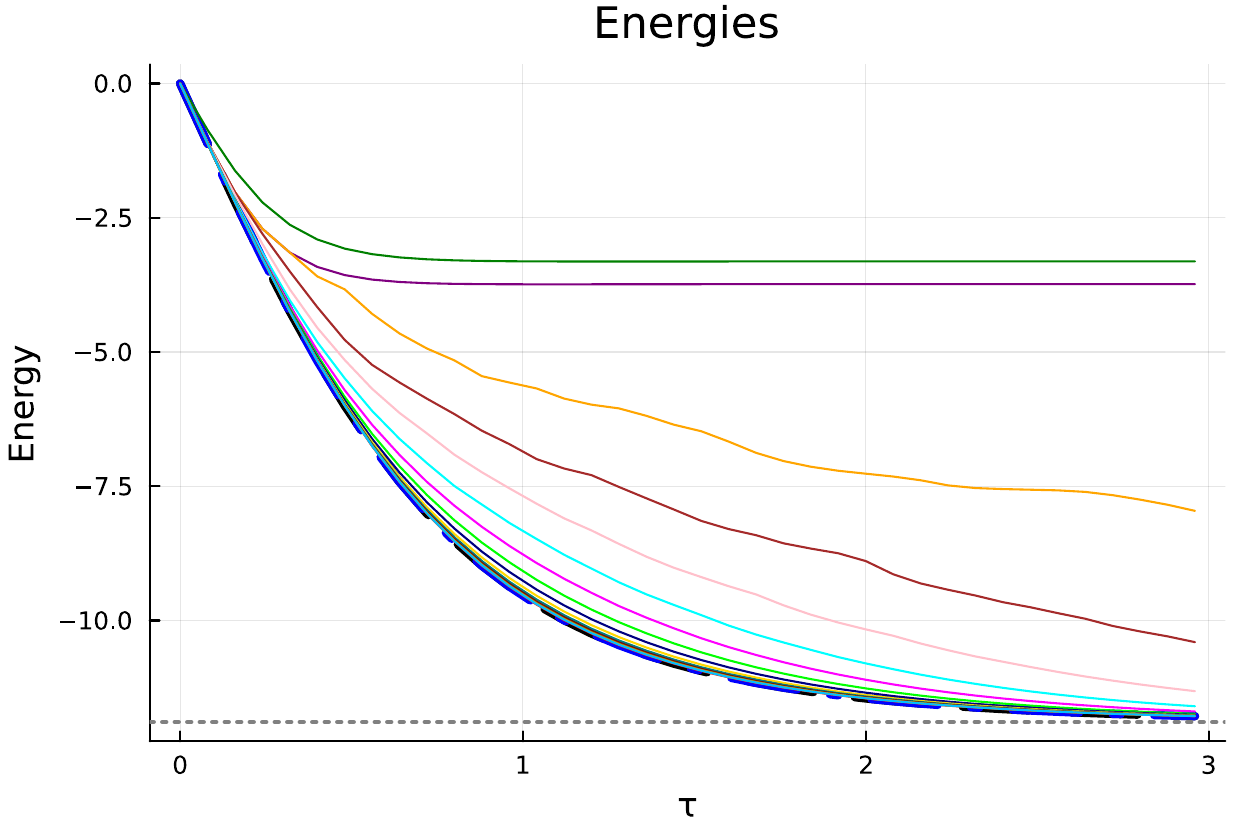}
        
    \end{subfigure}
    \hfill
    \begin{subfigure}[t]{0.48\textwidth}
        \centering
        \includegraphics[width=\linewidth]{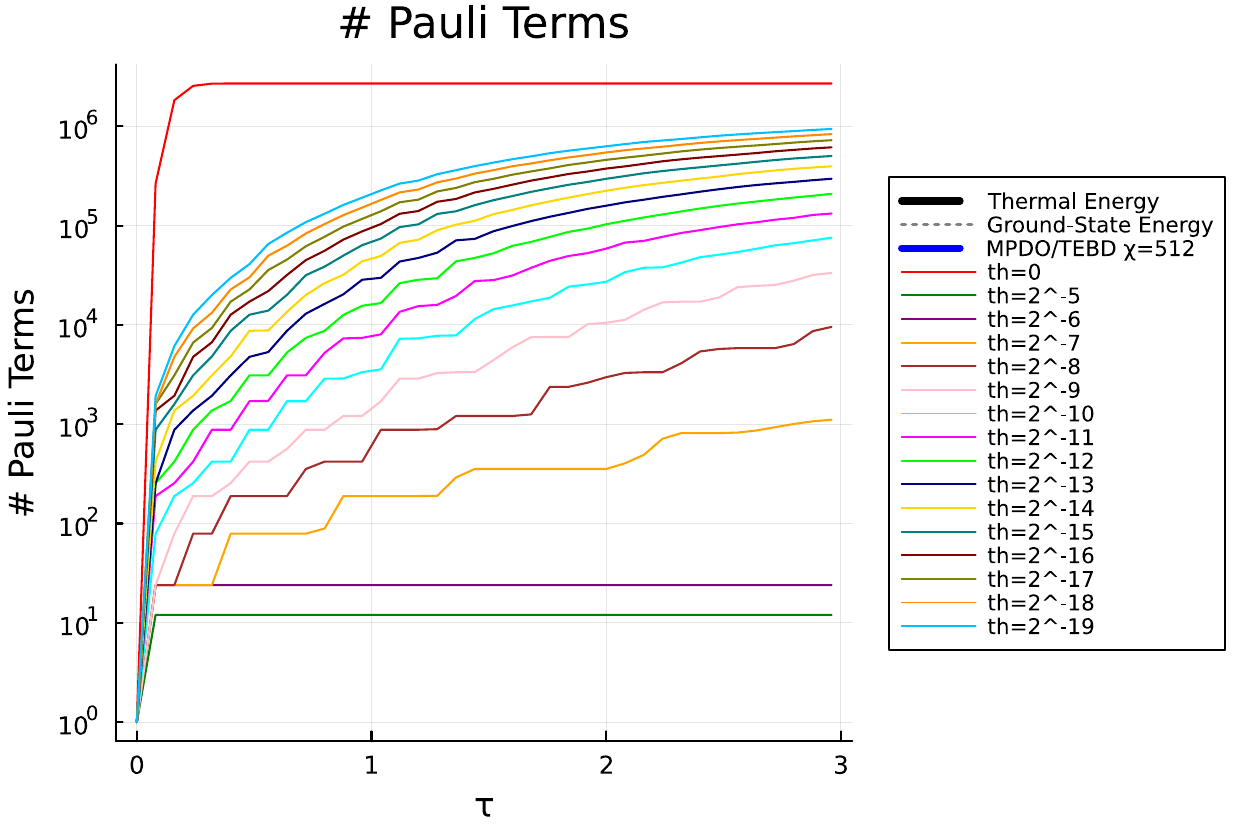}
        
    \end{subfigure}

    \vspace{0.3cm}

    \begin{subfigure}[t]{0.48\textwidth}
        \centering
        \includegraphics[width=\linewidth]{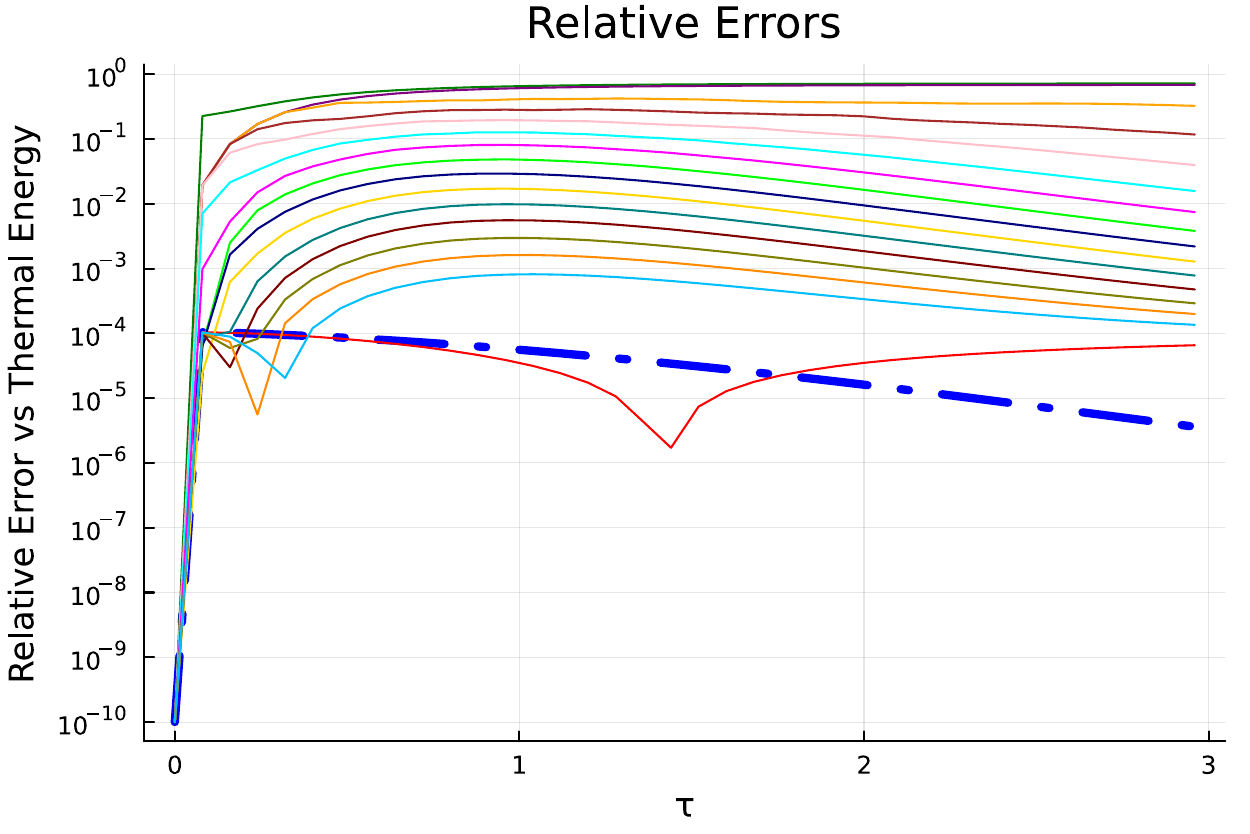}
        
    \end{subfigure}
    \hfill
    \begin{subfigure}[t]{0.48\textwidth}
        \centering
        \includegraphics[width=\linewidth]{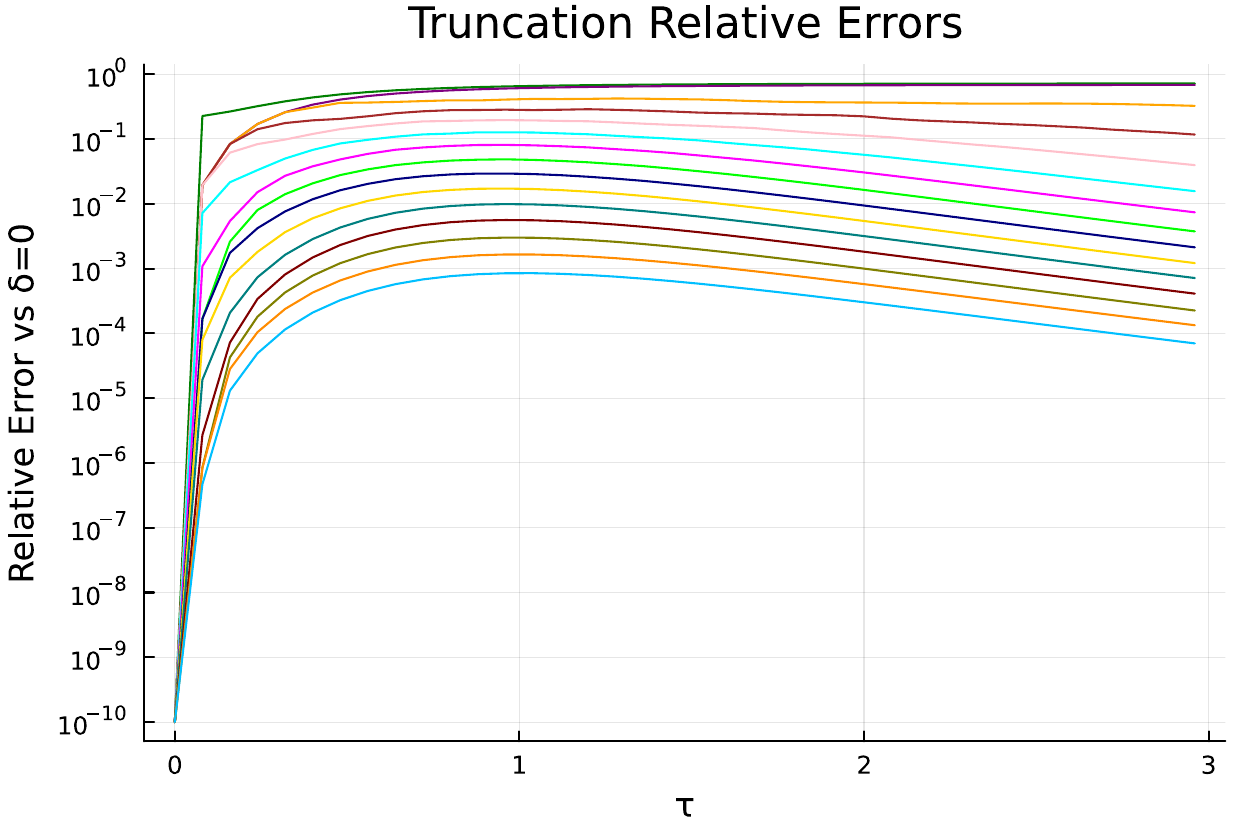}
        
    \end{subfigure}

        \caption{Finite-temperature performance of ITPP with coefficient-threshold truncation for the ordered-phase TFIM (\(J = 1\), \(h = 0.5\)) with \(N = 12\) spins, Trotter step \(\Delta \tau = 0.04\), and maximum imaginary time \(\tau = 3.0\).
    The panels show the thermal energy during imaginary-time evolution (top left), the number of retained Pauli terms (top right), the relative error with respect to the exact thermal energy (bottom left), and the relative error with respect to the untruncated ITPP result, \(\delta = 0\), which isolates the effect of truncation (bottom right).
    Dashed black lines indicate the exact Bogoliubov--de\,Gennes thermal energy where applicable.}
    
    \label{fig:PP_ITE_12_THERMAL}
\end{figure*}


\section{Numerical experiments}
\label{sec:experiments}

In this section, we present a systematic numerical study of imaginary-time Pauli propagation (ITPP) applied to the one-dimensional transverse-field Ising model (TFIM) with open boundary conditions. The Hamiltonian for a chain of \(N\) spins is
\begin{equation}
    H_{\mathrm{TFIM}}
    = - J \sum_{i=1}^{N-1} Z_i Z_{i+1}
      - h \sum_{i=1}^{N} X_i ,
    \label{eq:TFIM}
\end{equation}
where \(J\) denotes the nearest-neighbor coupling and \(h\) the transverse-field strength. As a reference, we benchmark our results against exact solutions obtained from the Bogoliubov--de\,Gennes (BdG) free-fermion formulation of the TFIM with open boundaries~\cite{Mbeng_2024}.

All simulations are performed using the \texttt{PauliPropagation.jl} Julia package~\cite{rudolph2025pauli}, extended to incorporate the imaginary-time propagation rules introduced in Sec.~\ref{Sec:ITPP}. A related Julia package, \texttt{PauliStrings.jl}~\cite{SciPostPhysCodeb.54}, also provides tools for working with Pauli-string representations.

Our numerical study is organized into three parts. We begin with a validation of the method, demonstrating that, in the absence of truncation, ITPP exactly reproduces Trotterized imaginary-time evolution and closely matches exact imaginary-time dynamics for small system sizes. This serves as a consistency check of the implementation and establishes a baseline for subsequent approximations.

We then turn to the primary focus of this work: the performance of ITPP for finite-temperature properties. We study the thermal energy for a fixed system size while varying the coefficient-threshold truncation parameter, quantifying the trade-off between accuracy and the number of retained Pauli terms as a function of inverse temperature \(\beta\). We then investigate scalability by analyzing how the number of Pauli terms required to achieve a fixed relative error depends on the system size. In addition to comparisons with exact BdG results, we benchmark ITPP against MPDO simulations based on TEBD, providing an objective comparison with established tensor-network methods.

Finally, we examine the low-temperature limit corresponding to ground-state preparation. For fixed system size, we again vary the coefficient-threshold truncation parameter to quantify the trade-off between accuracy and the number of retained Pauli terms as the imaginary time increases. To probe scalability in this regime, we adopt a complementary approach and study the behavior of the relative error as a function of system size for a fixed number of retained Pauli terms \(K\). This choice reflects the rapid growth of Pauli support at low temperatures, which makes maintaining a fixed accuracy increasingly costly. The resulting analysis illustrates that a fixed computational budget quickly becomes insufficient as the system size increases, highlighting the onset of a breakdown of the approximation at large imaginary times. We relate this behavior to the growth of Pauli support discussed in Sec.~\ref{Sec:ITPP}.

Throughout this section, we focus on the ordered phase of the TFIM (\(J > h\)), using the representative parameter values \(J = 1.0\) and \(h = 0.5\).

\subsection{Validation of imaginary-time Pauli propagation.}

We first verify that, in the absence of truncation, ITPP faithfully reproduces Trotterized imaginary-time evolution and compare its performance with exact imaginary-time evolution obtained via exact diagonalization. Simulations are performed for a chain of \(N = 10\) spins with Trotter step size \(\Delta \tau = 0.04\), truncation threshold \(\delta = 0\), and total imaginary time \(\tau = 20\). Throughout the evolution, we monitor the energy and its relative error with respect to the exact ground-state energy, defined as
\begin{equation}
\varepsilon(\tau) =
\frac{\left| E(\tau) - E_0 \right|}{\left| E_0 \right|},
\end{equation}
where \(E(\tau)\) is the energy at imaginary time \(\tau\), and \(E_0\) denotes the exact ground-state energy. The corresponding results are presented in Fig.~\ref{fig:ITE}.

The results show that ITPP exactly reproduces the Trotterized imaginary-time evolution, as expected, and closely follows the exact imaginary-time dynamics up to the onset of Trotter errors.

\subsection{Finite-temperature performance}

We next assess the finite-temperature performance of ITPP using coefficient-threshold truncation. Simulations are performed for a chain of \(N = 12\) spins with Trotter step size \(\Delta \tau = 0.04\), using truncation thresholds from \(2^{-19}\) to \(2^{-5}\), together with the idealized case of no truncation. The evolution is carried out up to inverse temperature \(\beta = 3\), focusing on the high-temperature regime where the Pauli representation is expected to remain relatively sparse.

We monitor the thermal energy, its relative error with respect to the exact result, the number of Pauli terms retained during the evolution, and the relative error with respect to the untruncated ITPP evolution (\(\delta = 0\)), which isolates the effect of truncation. As an additional benchmark, we compare the ITPP results with MPDO simulations based on TEBD, using bond dimension \(\chi = 512\). The corresponding results are shown in Fig.~\ref{fig:PP_ITE_12_THERMAL}.

The numerical results exhibit the expected behavior of ITPP under coefficient-threshold truncation. As the truncation threshold is decreased, an increasing number of Pauli terms is retained during the evolution, leading to progressively more accurate estimates of the thermal energy. This behavior is consistent with previous observations for real-time Pauli propagation.


\begin{figure*}
    \centering
    \textbf{Finite-temperature scalability of ITPP} \vspace{0.2cm}

    \begin{subfigure}[t]{0.32\textwidth}
        \centering
        \includegraphics[width=\linewidth]{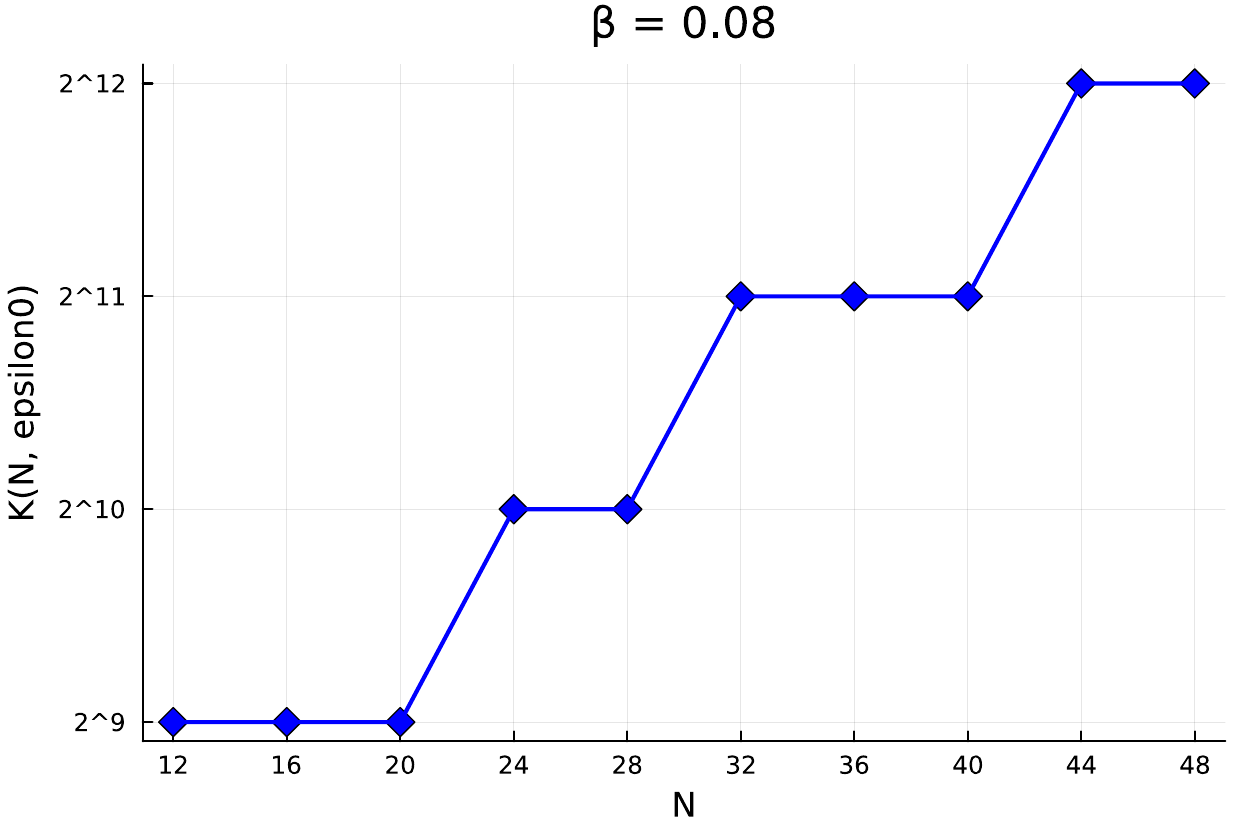}
     
    \end{subfigure}
    \hfill
    \begin{subfigure}[t]{0.32\textwidth}
        \centering
        \includegraphics[width=\linewidth]{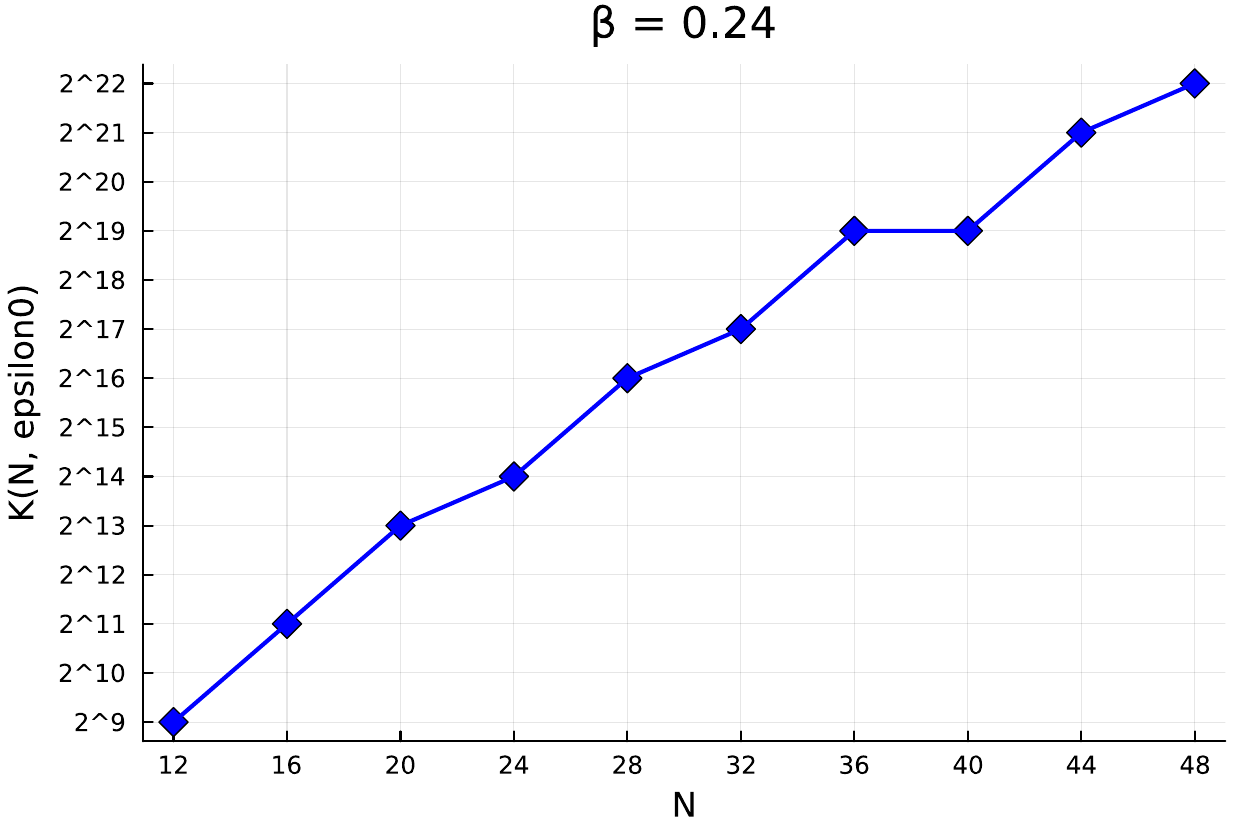}
       
    \end{subfigure}
    \hfill
    \begin{subfigure}[t]{0.32\textwidth}
        \centering
        \includegraphics[width=\linewidth]{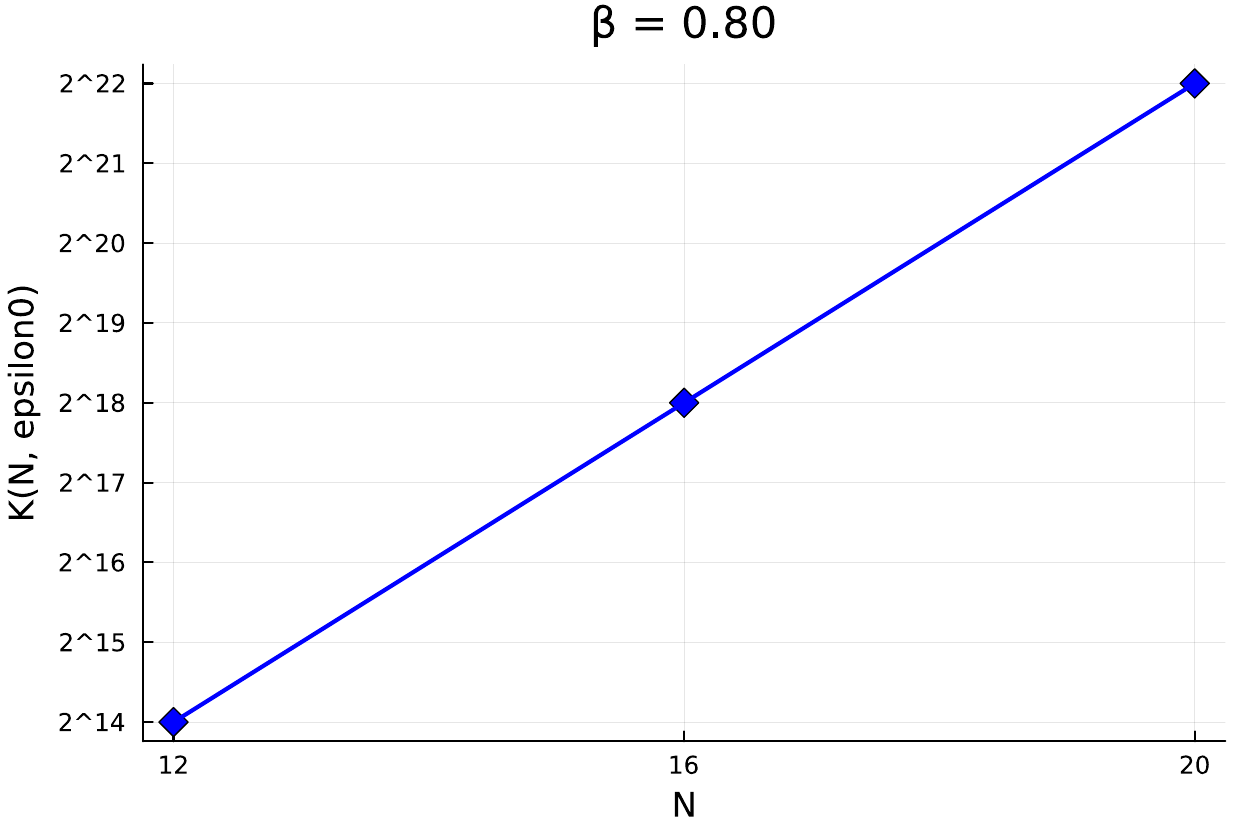}
       
    \end{subfigure}

    \vspace{0.3cm}

    \begin{subfigure}[t]{0.32\textwidth}
        \centering
        \includegraphics[width=\linewidth]{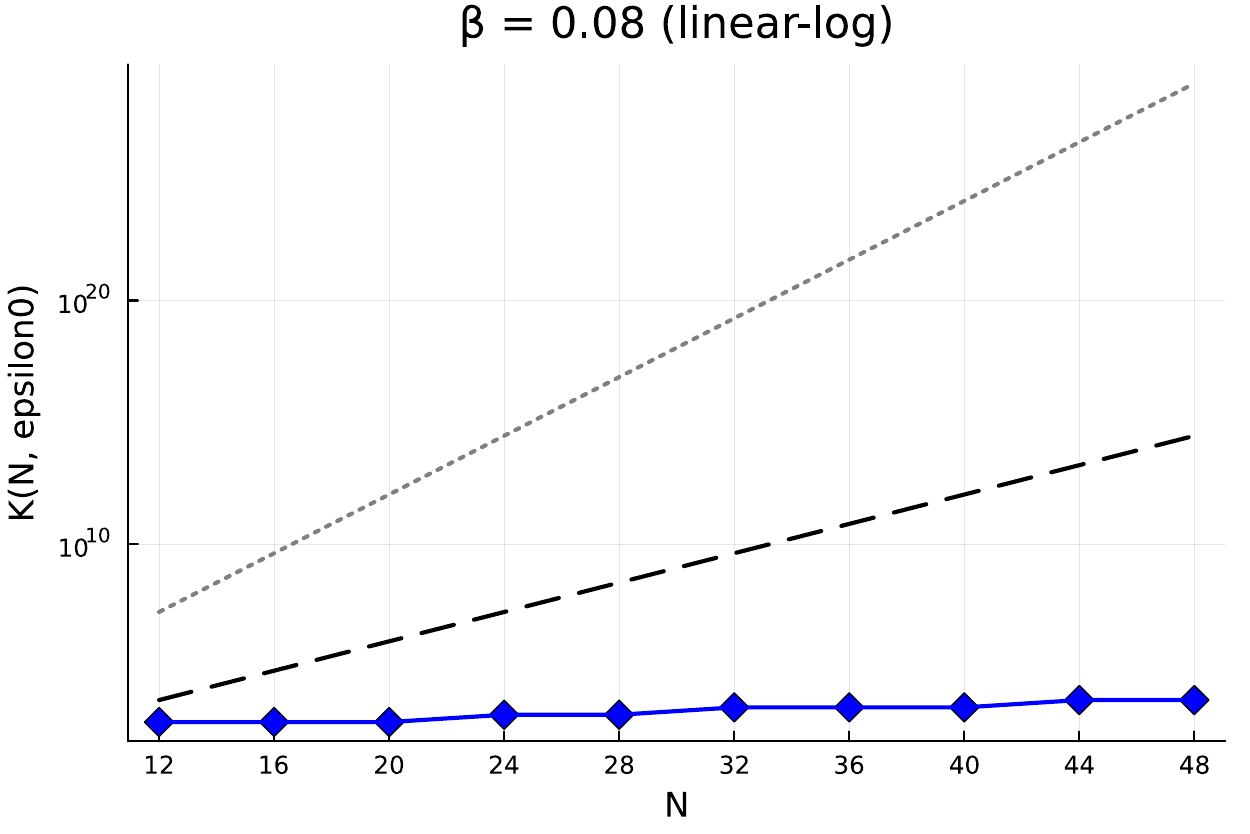}
        
    \end{subfigure}
    \hfill
    \begin{subfigure}[t]{0.32\textwidth}
        \centering
        \includegraphics[width=\linewidth]{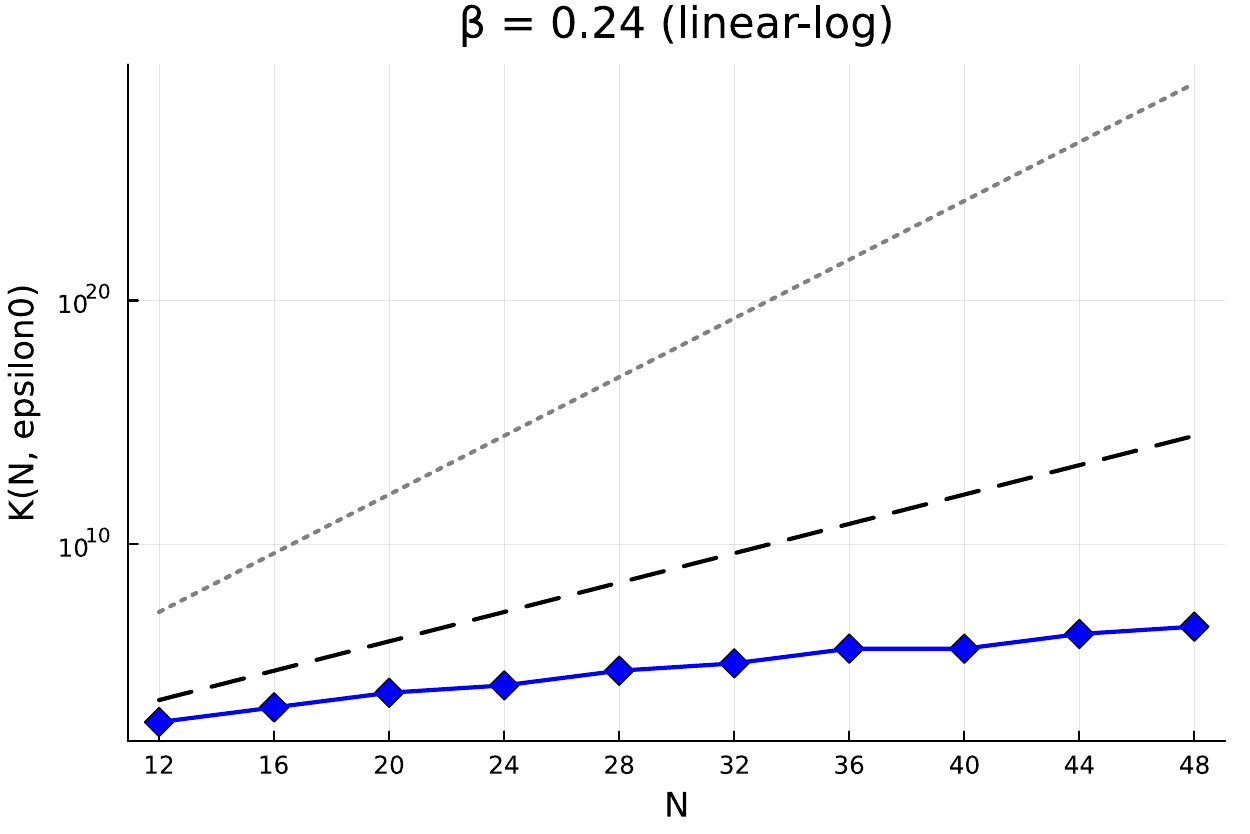}
       
    \end{subfigure}
    \hfill
    \begin{subfigure}[t]{0.32\textwidth}
        \centering
        \includegraphics[width=\linewidth]{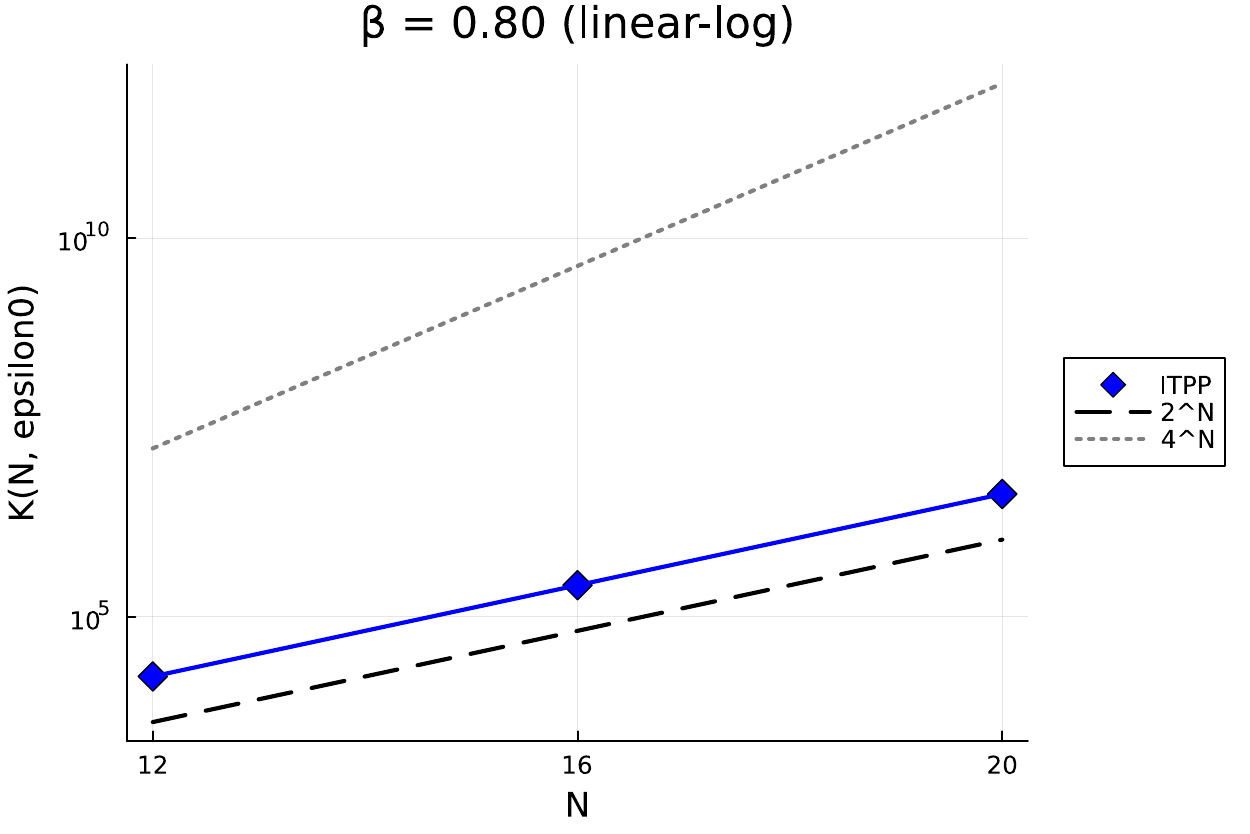}
        
    \end{subfigure}

    \caption{Finite-temperature scalability of ITPP with fixed-\(K\) truncation for the ordered-phase TFIM (\(J=1\), \(h=0.5\)). 
    The plotted quantity is \(K(N,\varepsilon_0)\), defined as the smallest number of Pauli terms required to achieve relative error \(\varepsilon \leq \varepsilon_0\) in the thermal energy, with \(\varepsilon_0 = 10^{-2}\).
    Columns correspond to inverse temperatures \(\beta = 0.08\), \(\beta = 0.24\), and \(\beta = 0.8\). 
    The top row shows the scaling on a linear scale, while the bottom row presents the same data on a semi-log scale, allowing direct comparison with exponential reference scalings proportional to \(2^N\) and \(4^N\).}
    
    \label{fig:thermal_scalability}
\end{figure*}


At short imaginary times, corresponding to high-temperature thermal states, accurate results can be obtained while retaining only a small fraction of the total number of Pauli terms. This indicates that the thermal state admits a sparse representation in the Pauli basis in this regime. By contrast, the untruncated evolution rapidly saturates to a large number of terms, reflecting the full growth of Pauli support.

As the imaginary time increases and the system approaches lower temperatures, an increasing number of Pauli terms acquire significant weight in the expansion, leading to a rapid growth in the number of terms required to represent the thermal state more accurately. In this regime, the Pauli expansion becomes increasingly dense, and the number of retained terms under truncation approaches that of the untruncated evolution. This behavior is consistent with the intuition discussed in Sec.~\ref{Sec:ITPP}, namely that sparsity is gradually lost as the system approaches the ground-state limit.

Coefficient-threshold truncation provides a flexible mechanism to balance accuracy and computational cost. LarAt short imaginary times, corresponding to high-temperature thermal states, accurate results can be obtained while retaining only a small fraction of the total number of Pauli terms. This indicates that the thermal state admits a sparse representation in the Pauli basis in this regime. By contrast, the untruncated evolution rapidly saturates to a large number of terms, reflecting the full growth of Pauli support.

As the imaginary time increases and the system approaches lower temperatures, an increasing number of Pauli terms acquire significant weight in the expansion, leading to a rapid growth in the number of terms required to represent the thermal state more accurately. In this regime, the Pauli expansion becomes increasingly dense, and the number of retained terms under truncation approaches that of the untruncated evolution. This behavior is consistent with the intuition discussed in Sec.~\ref{Sec:ITPP}, namely that sparsity is gradually lost as the system approaches the ground-state limit.
ger thresholds result in fewer retained terms and reduced computational effort, at the expense of accuracy, while smaller thresholds systematically improve the approximation.

Comparing with tensor-network methods, we observe that for short imaginary times ITPP achieves a relative error comparable to that obtained with MPDO simulations based on TEBD. As the imaginary time increases, however, the performance of ITPP deteriorates relative to MPDO, reflecting the increasing complexity of representing low-temperature states in the Pauli basis. This highlights that ITPP is most effective in the high-temperature regime, where the Pauli representation remains sparse.

Finally, to isolate the effect of truncation, we consider the relative error with respect to the untruncated ITPP evolution (\(\delta = 0\)). The results show that, at short imaginary times, small truncation thresholds yield very low errors while retaining only a small fraction of the Pauli terms. This confirms that accurate finite-temperature approximations can be obtained with substantially reduced computational cost in the high-temperature regime.

To assess the scalability of ITPP, we estimate the number of Pauli terms required to achieve a fixed target accuracy as a function of system size. Specifically, we consider the quantity \(K(N,\varepsilon_0)\), defined as the minimum number of retained Pauli terms needed to obtain a relative error \(\varepsilon \leq \varepsilon_0\) in the thermal energy. In our simulations, we fix \(\varepsilon_0 = 10^{-2}\) and evaluate the relative error with respect to the exact thermal energy.

To directly control the computational cost, we employ a fixed-\(K\) truncation strategy, in which the number of retained Pauli terms is capped at a maximum value \(K\). For each system size \(N\), we scan over values of \(K\) in the range \(2^9\) to \(2^{24}\), and compute the corresponding relative errors \(\varepsilon(N,K)\). The value \(K(N,\varepsilon_0)\) is then obtained as the smallest \(K\) for which the target accuracy is achieved. We carry out this procedure for system sizes ranging from \(N = 12\) to \(N = 48\) in steps of four.

The resulting scaling of \(K(N,\varepsilon_0)\) is shown in Fig.~\ref{fig:thermal_scalability} for three representative inverse temperatures: \(\beta = 0.08\) and \(\beta = 0.24\), corresponding to high-temperature regimes, and \(\beta = 0.8\), corresponding to a lower-temperature regime. These results provide a direct characterization of how the computational cost required to maintain a fixed accuracy grows with system size across different temperature regimes.

The scalability results reveal a strong dependence on the temperature regime. 
For high temperatures (\(\beta = 0.08\) and \(\beta = 0.24\)), the number of Pauli terms required to reach the target accuracy, \(K(N,\varepsilon_0)\), increases slowly over the range of system sizes considered. This behavior is particularly clear in the semi-log plots, where the ITPP data remain far below the reference scalings proportional to \(2^N\) and \(4^N\). Although these finite-size results do not determine the asymptotic scaling and do not rule out an exponential dependence with a smaller exponent, they indicate that the Pauli representation remains highly compressed in the high-temperature regime.

Comparing the two high-temperature cases, increasing \(\beta\) from \(0.08\) to \(0.24\) leads to a noticeable increase in \(K(N,\varepsilon_0)\), consistent with a gradual reduction in Pauli-basis sparsity. Nevertheless, the growth remains modest for the system sizes studied, suggesting favorable scaling behavior in the high-temperature regime considered here.

In contrast, for lower temperature (\(\beta = 0.8\)), the scaling behavior changes significantly. The semi-log plot shows a much steeper trend, approaching linear behavior, which indicates an approximately exponential growth of \(K(N,\varepsilon_0)\). Correspondingly, we are only able to extract data up to \(N=20\), as the required number of Pauli terms rapidly exceeds the accessible computational budget. This behavior reflects the loss of sparsity in the Pauli expansion as the system approaches lower temperatures.

Overall, these results support the sparsity intuition underlying ITPP. In the high-temperature regime, the thermal states remain well approximated by a sparse Pauli decomposition, allowing accurate estimates of the thermal energy with a relatively small number of retained terms. As the temperature decreases, more Pauli terms acquire significant weight and the computational cost grows rapidly. This suggests that ITPP is most useful in high-temperature regimes where the Pauli representation remains compressed.

\subsection{Low-temperature limit and ground-state behavior}

\begin{figure*}
    \centering
    \textbf{Ground-state performance of ITPP for \(N=12\)} \vspace{0.2cm}

    \begin{subfigure}[t]{0.48\textwidth}
        \centering
        \includegraphics[width=\linewidth]{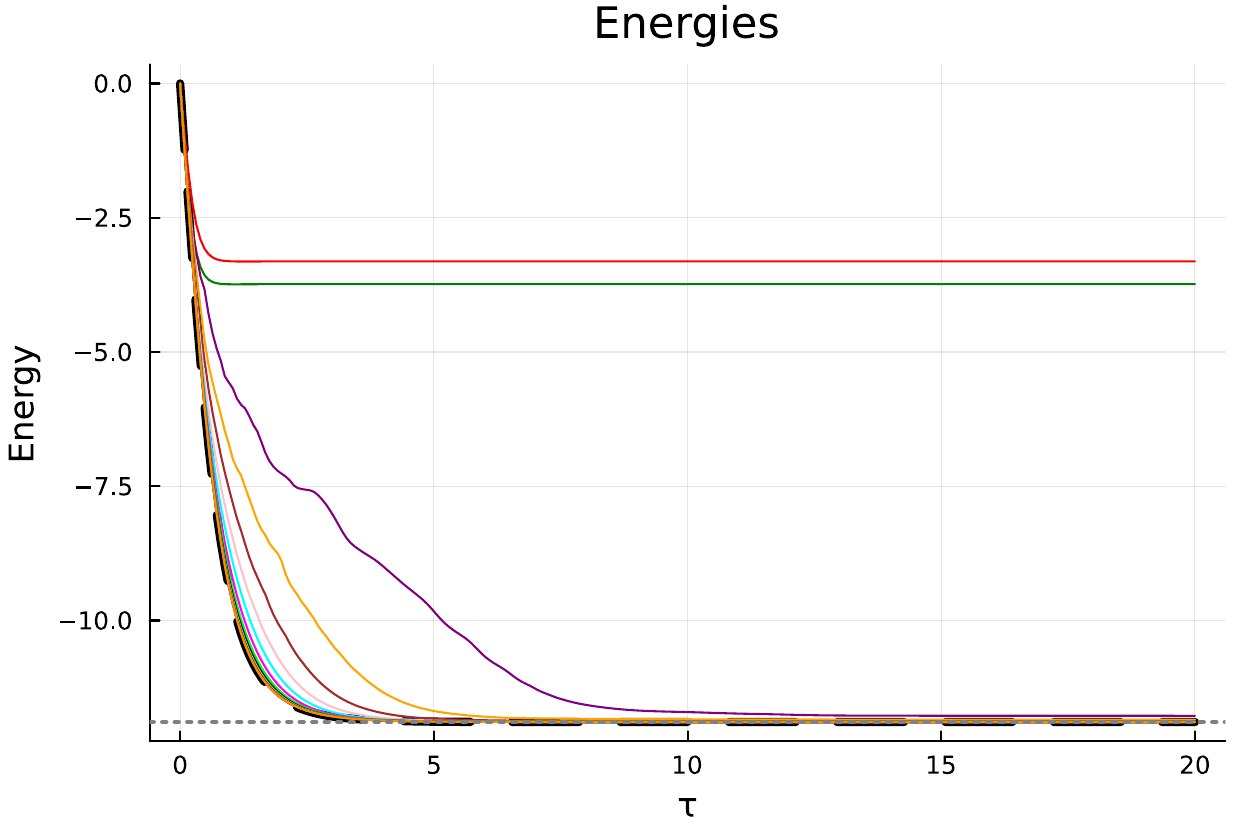}
    \end{subfigure}
    \hfill
    \begin{subfigure}[t]{0.48\textwidth}
        \centering
        \includegraphics[width=\linewidth]{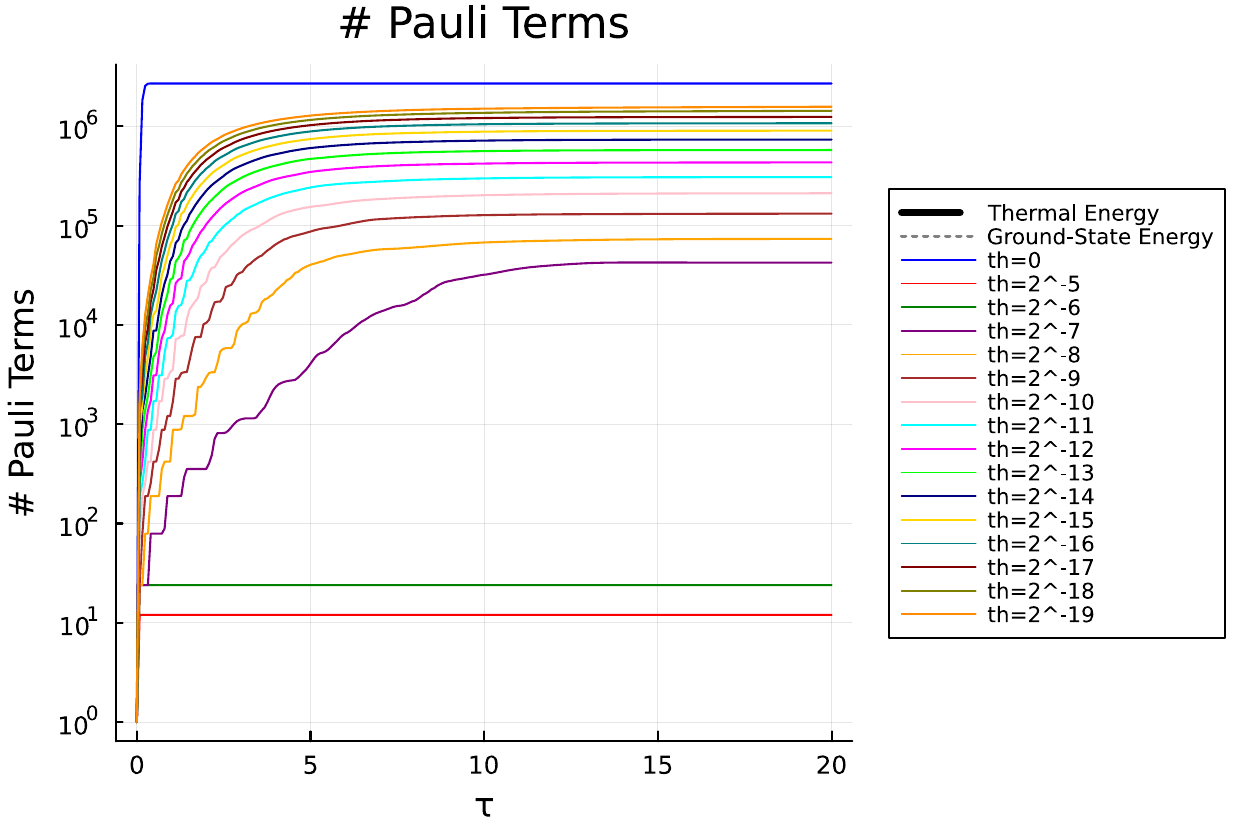}
    \end{subfigure}

    \vspace{0.3cm}

    \begin{subfigure}[t]{0.48\textwidth}
        \centering
        \includegraphics[width=\linewidth]{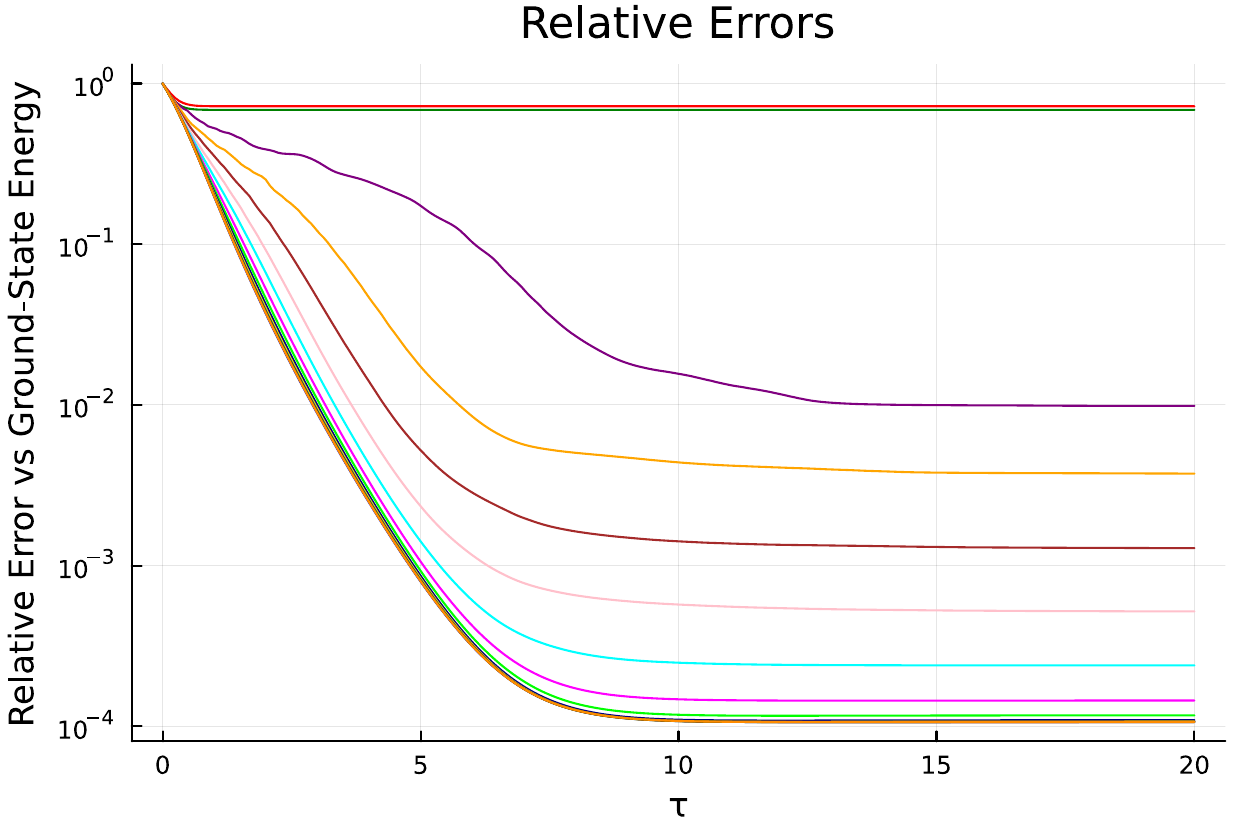}
    \end{subfigure}
    \hfill
    \begin{subfigure}[t]{0.48\textwidth}
        \centering
        \includegraphics[width=\linewidth]{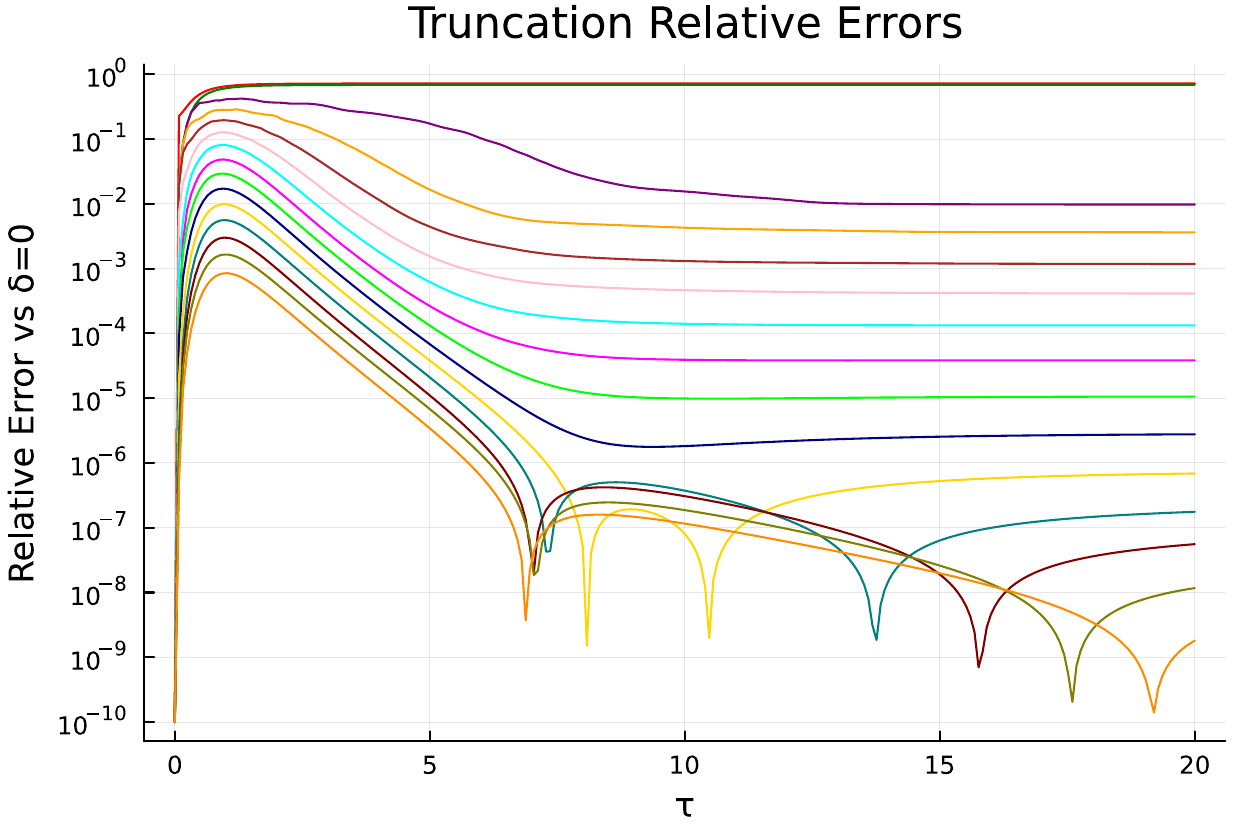}
    \end{subfigure}

    \caption{Low-temperature and ground-state performance of ITPP with coefficient-threshold truncation for the ordered-phase TFIM (\(J = 1\), \(h = 0.5\)) with \(N = 12\) spins, Trotter step \(\Delta \tau = 0.04\), and maximum imaginary time \(\tau = 20.0\).
    The panels show the energy during imaginary-time evolution (top left), the number of retained Pauli terms (top right), the relative error with respect to the exact ground-state energy (bottom left), and the relative error with respect to the untruncated ITPP result, \(\delta = 0\), which isolates the effect of truncation (bottom right).
    Dashed black lines indicate the exact Bogoliubov--de\,Gennes ground-state energy where applicable.}
    
    \label{fig:PP_ITE_12_GS}
\end{figure*}

We next assess the performance of ITPP in the low-temperature limit corresponding to ground-state estimation of the ordered-phase TFIM (\(J = 1\), \(h = 0.5\)). As in the finite-temperature case, we employ a coefficient-threshold truncation strategy with thresholds ranging from \(2^{-19}\) to \(2^{-5}\), together with the untruncated case \(\delta = 0\). Simulations are performed for a chain of \(N = 12\) spins with Trotter step size \(\Delta \tau = 0.04\), and the imaginary-time evolution is extended up to \(\tau = 20\). Throughout the evolution, we track the energy, the relative error with respect to the exact ground-state energy, the number of retained Pauli terms, and the relative error with respect to the untruncated ITPP evolution, which isolates the effect of truncation. The results are shown in Fig.~\ref{fig:PP_ITE_12_GS}.

The results follow the expected trend under coefficient-threshold truncation: decreasing the threshold leads to a larger retained Pauli support and correspondingly improved accuracy. This behavior is consistent with that observed in the finite-temperature regime and in real-time Pauli propagation.

A key question is how large a Pauli basis is required to obtain a reliable approximation of the ground-state energy. From Fig.~\ref{fig:PP_ITE_12_GS}, we observe that significant improvements in accuracy occur for thresholds below approximately \(2^{-7}\). For instance, at \(\delta = 2^{-7}\), ITPP achieves a relative error of order \(10^{-2}\) while retaining \(42{,}466\) Pauli terms, compared to \(2{,}704{,}156\) terms in the untruncated evolution. This indicates that, even in this regime, substantial reductions in computational cost are possible while maintaining acceptable accuracy.

Across the range of thresholds, we observe a clear trade-off between accuracy and computational cost. Lower thresholds yield improved approximations at the expense of a rapidly increasing number of retained terms, while larger thresholds lead to a significant loss of accuracy. Compared to the finite-temperature case, however, this trade-off becomes considerably more demanding, as the number of relevant Pauli terms grows strongly with imaginary time. This is reflected in the convergence of the truncated dynamics toward the untruncated case as \(\tau\) increases.


For sufficiently large thresholds, the number of retained Pauli terms drops sharply, resulting in poor approximations of the ground state. This can be understood in terms of the branching structure of Pauli propagation: discarding small coefficients at early stages removes contributions that would later become important, thereby restricting the accessible operator space.

The relative error with respect to the untruncated evolution further clarifies this behavior. In contrast with the high-temperature regime, achieving low error at large imaginary times requires retaining a substantial fraction of the Pauli terms. This reflects the increasing importance of a large number of Pauli strings as the system approaches the ground state.

Overall, these results align with the intuition underlying ITPP: while truncation provides a controllable approximation, the rapid growth of Pauli support at low temperatures leads to a substantial increase in computational cost. Ground-state estimation therefore represents a significantly more challenging regime, although intermediate thresholds can still offer useful approximations with reduced resources.

To further assess scalability in the low-temperature regime, we use the fixed-\(K\) truncation scheme introduced above and study the relative error as a function of system size under a fixed computational budget.

\begin{figure*}
    \centering
    \textbf{ITPP with fixed-\(K\) truncation and increasing \(N\)} \vspace{0.2cm} \\
    \includegraphics[scale=0.32]{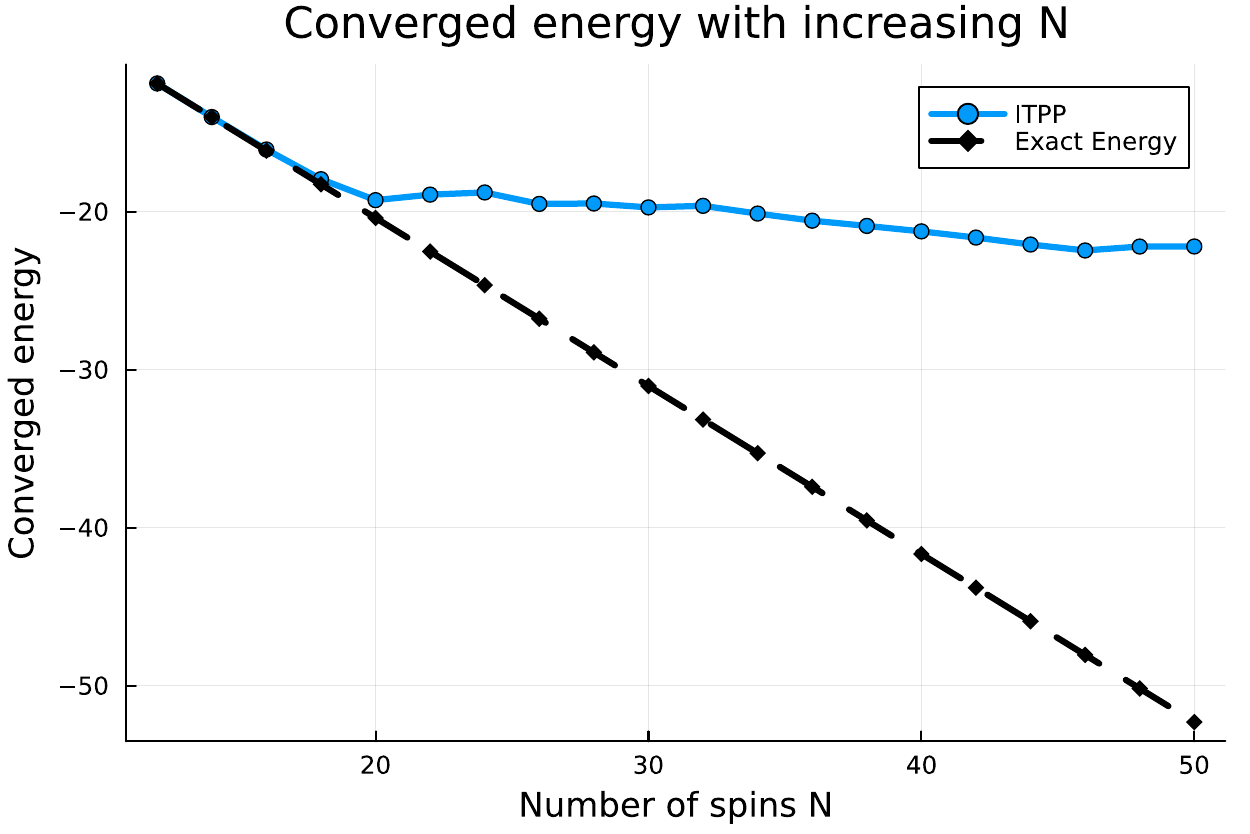}
    \includegraphics[scale=0.32]{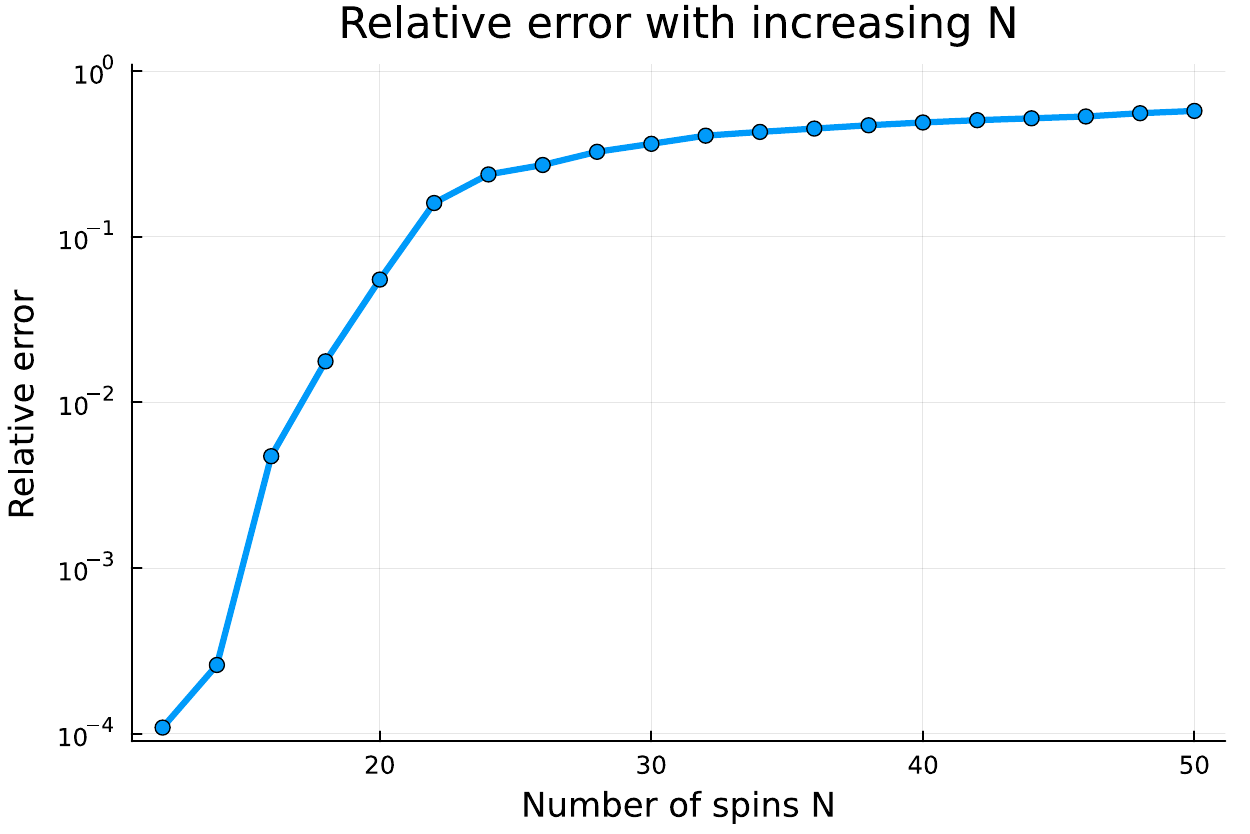}
    \caption{Scalability of ITPP with fixed-\(K\)
truncation for the one-dimensional TFIM in the ordered phase
(\(J = 1\), \(h = 0.5\)), for system sizes ranging from \(N = 12\) to
\(N = 50\) spins in steps of two.
    The left panel shows the converged ground-state energy obtained with ITPP as a function of system size \(N\), while the right panel displays the corresponding relative error (on a logarithmic scale) with respect to the exact ground-state energy computed using the Bogoliubov--de\,Gennes solution.
    The truncation parameter is fixed to \(K = 2{,}704{,}156\), equal to the number of Pauli terms generated for \(N = 12\) in the absence of truncation.
    The results indicate the range of system sizes for which this fixed Pauli basis size is sufficient to accurately capture the underlying operator growth under imaginary-time evolution, as well as the onset of a breakdown when the fixed computational budget becomes inadequate.}
    \label{fig:PP_ITE_fixedK}
\end{figure*}

Specifically, we set \(K\) equal to the number of Pauli terms generated for \(N = 12\) in the untruncated evolution, namely \(K = 2{,}704{,}156\). Keeping this value fixed, we simulate chains with system sizes ranging from \(N = 12\) to \(N = 50\) in steps of two. The imaginary-time evolution is carried out up to \(\tau = 20\), and the resulting energies are compared with exact ground-state values. The results are shown in Fig.~\ref{fig:PP_ITE_fixedK}.


We observe that fixed-\(K\) truncation yields accurate ground-state energies up to system sizes of approximately \(N \simeq 20\). Beyond this point, the relative error increases rapidly, and the estimated energy ceases to improve with increasing system size, indicating a breakdown of the approximation due to the fixed computational budget. This behavior reflects the rapid growth of Pauli support at large imaginary times: as the system size increases, an increasing number of Pauli terms is required to accurately represent the evolving state.

The onset of this breakdown occurs when the fixed value of \(K\) becomes insufficient to capture the relevant operator space. In particular, the loss of accuracy coincides with the regime where \(K\) becomes small compared to \(2^N\), suggesting that the number of Pauli terms required to approximate the ground-state energy grows rapidly with system size. At the same time, the results show that accurate approximations can still be achieved for moderate system sizes while retaining a number of terms that is significantly smaller than the full Pauli basis.

These observations should be interpreted with caution. Since the truncation parameter \(K\) is fixed to a single value determined from the \(N = 12\) system, the present results do not allow for a definitive characterization of the asymptotic scaling of the algorithm. Nevertheless, they provide clear evidence that maintaining a fixed computational budget becomes increasingly challenging in the low-temperature regime, consistent with the rapid growth of Pauli support discussed in Sec.~\ref{Sec:ITPP}.

\section{Conclusions}
\label{sec:conclusion}

In this work, we extended the Pauli Propagation framework to imaginary-time evolution. By deriving explicit update rules for the propagation of Pauli strings under imaginary-time operators generated by Pauli strings, we introduced imaginary-time Pauli Propagation (ITPP), an operator-based algorithm for approximating thermal and ground-state properties in the Pauli basis. In the absence of truncation, ITPP exactly reproduces Trotterized imaginary-time evolution, providing a consistency check of the formalism and implementation.

We benchmarked ITPP on the one-dimensional transverse-field Ising model in the ordered phase. Our numerical results show that the method is most effective in the finite-temperature regime, where the density matrix remains well approximated by a sparse Pauli expansion. In this regime, coefficient-threshold truncation yields accurate thermal energies while retaining only a small fraction of the Pauli terms generated by untruncated evolution. Comparisons with exact results and MPDO simulations based on TEBD show that ITPP achieves competitive accuracy at high temperatures, while offering a distinct Pauli-basis representation of the thermal state.

We also investigated the scalability of ITPP by estimating the number of retained Pauli terms \(K(N,\varepsilon_0)\) required to achieve a fixed target accuracy in the thermal energy. For high-temperature states, this number grows slowly over the accessible system sizes and remains far below the reference scalings \(2^N\) and \(4^N\). As the temperature is lowered, however, the required Pauli support increases rapidly. This behavior supports the central intuition behind the method: ITPP is effective when the thermal state admits a compressed Pauli-basis approximation in the high-temperature limit, but becomes increasingly costly as the state approaches the ground-state limit.

The ground-state experiments further highlight this limitation. Although coefficient-threshold truncation can still provide useful approximations with substantially fewer terms than untruncated evolution, maintaining accuracy at large imaginary times requires retaining a much larger fraction of the Pauli expansion. Fixed-\(K\) experiments show that a constant computational budget eventually becomes insufficient as the system size increases, leading to a breakdown of the approximation. These observations are consistent with the expected growth of Pauli support at low temperatures and with the intrinsic difficulty of ground-state preparation for general local Hamiltonians.

Overall, our results establish ITPP as a useful complementary framework for imaginary-time simulation, particularly in regimes where the density matrix admits a sparse Pauli-basis approximation. The method is not intended to replace tensor-network, Monte Carlo, or quantum imaginary-time approaches, but rather provides a distinct operator-space perspective with different trade-offs. In particular, ITPP does not rely on low-entanglement structure or stochastic sampling, and a single propagated density operator can be reused to evaluate multiple observables without rerunning the simulation. These features may be useful in settings where existing approaches face difficulties, such as highly connected or higher-dimensional systems, sign-problem instances, or as a classical benchmark for quantum imaginary-time algorithms.

Several directions remain open for future work. On the numerical side, an important next step is to test ITPP on two- and three-dimensional many-body systems, where tensor-network methods are expected to become more challenging due to increased connectivity and entanglement structure. Such experiments would help clarify whether Pauli-space sparsity can provide advantages in regimes that are difficult for conventional tensor-network simulations. It would also be valuable to explore more adaptive truncation strategies, improved Pauli-based ansätze, and higher-order imaginary-time product formulas.

On the theoretical side, a particularly interesting question is whether the analytical success of real-time Pauli Propagation for random circuits with anti-concentration properties has an analogue in imaginary time. One possible direction is to study imaginary-time counterparts of random circuits, obtained by replacing unitary gates with corresponding non-unitary imaginary-time gates, and to ask whether such dynamics admit efficient sparse Pauli descriptions. Developing rigorous guarantees for weight-based truncation, or identifying ensembles and Hamiltonian classes for which ITPP is provably efficient, would substantially sharpen the theoretical understanding of the method.

Finally, the extension to imaginary-time evolution suggests a natural pathway toward simulating more general open quantum system dynamics when combined with real-time Pauli Propagation. One possible route is to work with vectorized Lindblad dynamics, where the generator contains both Hermitian, unitary contributions and anti-Hermitian, non-unitary contributions. In such a formulation, the unitary part could be treated using real-time Pauli Propagation, while the non-unitary part could be handled using the imaginary-time extension introduced here. Related strategies have been explored for simulating open quantum dynamics on quantum hardware~\cite{PRXQuantum.3.010320}, and vectorized Pauli-based approaches have been investigated in Ref.~\cite{rudolph2023classicalsurrogatesimulationquantum}. Developing this direction would require reformulating the present method in terms of vectorized operators, and we leave a detailed exploration to future work.


\section*{Acknowledgments}

RGL thanks Shao-Hen Chiew and Alberto Acevedo Meléndez for valuable discussions and feedbacks. This work was supported by the project PID2023-152724NA-I00, with funding from MCIU/AEI/10.13039/501100011033 and FSE+, the Severo Ochoa Grant CEX2023-001292-S, Generalitat Valenciana grant CIPROM/2022/66, the Ministry of Economic Affairs and Digital Transformation of the Spanish Government through the QUANTUM ENIA project call - QUANTUM SPAIN project, and by the European Union through the Recovery, Transformation and Resilience Plan - NextGenerationEU within the framework of the Digital Spain 2026 Agenda, and by the CSIC Interdisciplinary Thematic Platform (PTI+) on Quantum Technologies (PTI-QTEP+). This project has also received funding from the European Union’s Horizon 2020 research and innovation program under grant agreement CaLIGOLA MSCA-2021-SE-01-101086123. RGL is funded by grant CIACIF/2021/136 from Generalitat Valenciana.

\section*{Added Note}

After the first version of the present work was made publicly available on arXiv, Ref.~\cite{rudolph2026thermalstatesimulationpauli} appeared on arXiv. That work studies a closely related approach to imaginary-time Pauli propagation. In particular, it provides rigorous error bounds for weight-truncation schemes for first-order Trotterized imaginary-time evolution under locality assumptions, as well as bounds for weight- and small-angle-truncation schemes in qDRIFT-based stochastic product formulas. It also extends the framework to Majorana propagation, enabling applications to fermionic systems.

\bibliographystyle{apsrev4-2}
\bibliography{references}

\newpage

\onecolumngrid
\appendix
\section{Pauli Propagation Rules for Imaginary-Time Evolution}
\label{app:ITPP_propagation_rules}

In this appendix, we derive the propagation rules governing the imaginary-time evolution of a Pauli string \(P\) generated by another Pauli string \(Q\). We consider the non-unitary imaginary-time evolution operator
\begin{equation}
    V_Q(\tau) = e^{-\frac{\tau}{2} Q},
\end{equation}
where \(\tau \in \mathbb{R}\) denotes the imaginary-time and \(Q\) is a Pauli string satisfying \(Q^2=\mathbb{I}\). The imaginary-time–evolved Pauli operator is defined as
\begin{equation}
    P(\tau) = V_Q^\dagger(\tau)\, P \, V_Q(\tau).
    \label{eq:Pauli_ITE_def}
\end{equation}

Using the expansion
\begin{equation}
    V_Q(\tau)
    = \cosh\!\left(\frac{\tau}{2}\right)\mathbb{I}
    - \sinh\!\left(\frac{\tau}{2}\right) Q,
    \label{eq:ITE_expansion}
\end{equation}
and substituting it into Eq.~\eqref{eq:Pauli_ITE_def}, we obtain the general expression
\begin{equation}
    V_Q^\dagger(\tau)\, P \, V_Q(\tau)
    =
    \cosh^2\!\left(\frac{\tau}{2}\right) P
    - \cosh\!\left(\frac{\tau}{2}\right)\sinh\!\left(\frac{\tau}{2}\right)
    (QP + PQ)
    + \sinh^2\!\left(\frac{\tau}{2}\right) Q P Q .
    \label{eq:ITE_rules_general}
\end{equation}

Since \(P\) and \(Q\) are Pauli strings, they either commute or anticommute, leading to two distinct cases.

\subsubsection*{Anticommuting case \(\{P,Q\}=0\).}

If \(P\) and \(Q\) anticommute, the second term in Eq.~\eqref{eq:ITE_rules_general} vanishes because \(QP+PQ=0\). Using \(Q P Q = -P\), we find
\begin{equation}
    V_Q^\dagger(\tau)\, P \, V_Q(\tau)
    =
    \left[\cosh^2\!\left(\frac{\tau}{2}\right)
    - \sinh^2\!\left(\frac{\tau}{2}\right)\right] P
    = P,
\end{equation}
where we have used the identity \(\cosh^2(x)-\sinh^2(x)=1\).

\subsubsection*{Commuting case \([P,Q]=0\)}
If \(P\) and \(Q\) commute, \(QP = PQ\), and Eq.~\eqref{eq:ITE_rules_general} simplifies to
\begin{equation}
    V_Q^\dagger(\tau)\, P \, V_Q(\tau)
    =
    \left[\cosh^2\!\left(\frac{\tau}{2}\right)
    + \sinh^2\!\left(\frac{\tau}{2}\right)\right] P
    - 2 \cosh\!\left(\frac{\tau}{2}\right)\sinh\!\left(\frac{\tau}{2}\right) QP .
\end{equation}
Using the identities
\(\cosh^2(x)+\sinh^2(x)=\cosh(2x)\) and
\(2\cosh(x)\sinh(x)=\sinh(2x)\), this expression can be written as
\begin{equation}
    V_Q^\dagger(\tau)\, P \, V_Q(\tau)
    =
    \cosh(\tau)\, P - \sinh(\tau)\, QP .
\end{equation}

\subsubsection*{Summary}

The propagation of a Pauli string \(P\) under imaginary-time evolution generated by a Pauli string \(Q\) is therefore given by
\begin{equation}
    V_Q^\dagger(\tau)\, P \, V_Q(\tau)
    =
    \begin{cases}
        P, & \{P,Q\}=0, \\[6pt]
        \cosh(\tau)\, P - \sinh(\tau)\, QP, & [P,Q]=0 .
    \end{cases}
    \label{eq:ITE_Pauli_rules}
\end{equation}
These propagation rules form the basis of Pauli propagation under imaginary-time evolution employed in ITPP.

\section{First-Order Trotter Error Bound for Imaginary-Time Evolution}
\label{app:trotter_imag_time}

In this appendix, we derive a first-order Trotter error bound for the imaginary-time evolution operator associated with a Hamiltonian
\begin{equation}
    H = \sum_{j=1}^{L} \alpha_j h_j,
\end{equation}
where each \(h_j\) is a Pauli string and \(\alpha_j \in \mathbb{R}\). We consider the exact imaginary-time evolution operator
\begin{equation}
    V(\tau) = e^{-\frac{\tau}{2}H},
\end{equation}
and its first-order Trotter approximation
\begin{equation}
    V_{\mathrm{Trotter}}(\tau)
    =
    \left(
        \prod_{j=1}^{L}
        e^{-\frac{\Delta\tau}{2}\alpha_j h_j}
    \right)^n,
    \qquad
    \tau = n \Delta\tau.
\end{equation}
Our goal is to bound the error
\begin{equation}
    \bigl\|
        V_{\mathrm{Trotter}}(\tau) - V(\tau)
    \bigr\|.
\end{equation}

Throughout, we use the operator norm (spectral norm),
\begin{equation}
    \|X\| := \sup_{\|v\|_2 = 1} \|Xv\|_2,
\end{equation}
which satisfies the triangle inequality,
\begin{equation}
    \|X+Y\| \le \|X\| + \|Y\|,
\end{equation}
and submultiplicativity,
\begin{equation}
    \|XY\| \le \|X\|\,\|Y\|.
\end{equation}
We also use the standard bound
\begin{equation}
    \|e^X\| \le e^{\|X\|},
\end{equation}
valid for any bounded operator \(X\), and the commutator estimate
\begin{equation}
    \|[X,Y]\| = \|XY-YX\| \le 2\|X\|\,\|Y\|.
\end{equation}

Since each \(h_j\) is a Pauli string, we have
\begin{equation}
    h_j^\dagger = h_j,
    \qquad
    h_j^2 = I,
    \qquad
    \|h_j\| = 1.
\end{equation}

\subsection{Local error for a single Trotter step}

Define
\begin{equation}
    A_j := -\frac{\Delta\tau}{2}\alpha_j h_j.
\end{equation}
Then the exact and Trotterized one-step propagators are
\begin{equation}
    E(\Delta\tau) := e^{\sum_{j=1}^{L} A_j}
    = e^{-\frac{\Delta\tau}{2}H},
    \qquad
    S(\Delta\tau) := \prod_{j=1}^{L} e^{A_j}.
\end{equation}
We begin by bounding the local error
\begin{equation}
    \|S(\Delta\tau) - E(\Delta\tau)\|.
\end{equation}

To second order in \(\Delta\tau\), each exponential admits the expansion
\begin{equation}
    e^{A_j}
    =
    I + A_j + \frac{1}{2}A_j^2 + \mathcal{O}(\Delta\tau^3),
\end{equation}
where \(\|A_j\| = \mathcal{O}(\Delta\tau)\). Expanding the ordered product and keeping terms up to second order yields
\begin{equation}
    \prod_{j=1}^{L} e^{A_j}
    =
    I
    + \sum_{j} A_j
    + \frac{1}{2}\sum_{j} A_j^2
    + \sum_{j<k} A_j A_k
    + \mathcal{O}(\Delta\tau^3).
\end{equation}
On the other hand,
\begin{equation}
    e^{\sum_j A_j}
    =
    I
    + \sum_j A_j
    + \frac{1}{2}\left(\sum_j A_j\right)^2
    + \mathcal{O}(\Delta\tau^3),
\end{equation}
and since
\begin{equation}
    \left(\sum_j A_j\right)^2
    =
    \sum_j A_j^2
    +
    \sum_{j<k}(A_jA_k + A_kA_j),
\end{equation}
we obtain
\begin{equation}
    e^{\sum_j A_j}
    =
    I
    + \sum_j A_j
    + \frac{1}{2}\sum_j A_j^2
    + \frac{1}{2}\sum_{j<k}(A_jA_k + A_kA_j)
    + \mathcal{O}(\Delta\tau^3).
\end{equation}
Subtracting the two expansions gives
\begin{equation}
    S(\Delta\tau) - E(\Delta\tau)
    =
    \frac{1}{2}\sum_{j<k}(A_jA_k - A_kA_j)
    + \mathcal{O}(\Delta\tau^3),
\end{equation}
that is,
\begin{equation}
    S(\Delta\tau) - E(\Delta\tau)
    =
    \frac{1}{2}\sum_{j<k}[A_j,A_k]
    + \mathcal{O}(\Delta\tau^3).
\end{equation}
Taking norms and using the triangle inequality,
\begin{equation}
    \|S(\Delta\tau) - E(\Delta\tau)\|
    \le
    \frac{1}{2}\sum_{j<k}\|[A_j,A_k]\|
    + \mathcal{O}(\Delta\tau^3).
\end{equation}
Now,
\begin{equation}
    [A_j,A_k]
    =
    \left[-\frac{\Delta\tau}{2}\alpha_j h_j,\,
          -\frac{\Delta\tau}{2}\alpha_k h_k\right]
    =
    \frac{\Delta\tau^2}{4}\alpha_j\alpha_k [h_j,h_k],
\end{equation}
so that
\begin{equation}
    \|[A_j,A_k]\|
    =
    \frac{\Delta\tau^2}{4}
    |\alpha_j\alpha_k|\,\|[h_j,h_k]\|.
\end{equation}
Substituting this into the previous expression yields
\begin{equation}
    \|S(\Delta\tau) - E(\Delta\tau)\|
    \le
    \frac{\Delta\tau^2}{8}
    \sum_{j<k}
    |\alpha_j\alpha_k|\,\|[h_j,h_k]\|
    +
    \mathcal{O}(\Delta\tau^3).
\end{equation}
Defining
\begin{equation}
    C
    :=
    \sum_{j<k}
    |\alpha_j\alpha_k|\,\|[h_j,h_k]\|,
\end{equation}
we obtain the local bound
\begin{equation}
    \boxed{
    \|S(\Delta\tau) - E(\Delta\tau)\|
    \le
    \frac{\Delta\tau^2}{8}\, C
    + \mathcal{O}(\Delta\tau^3).
    }
\end{equation}

For Pauli strings, \(\|h_j\|=1\) and therefore
\begin{equation}
    \|[h_j,h_k]\|
    \le 2\|h_j\|\,\|h_k\|
    = 2,
\end{equation}
so \(C\) is finite and controlled entirely by the coefficients \(\alpha_j\) and the non-commutativity structure of the Hamiltonian.

\subsection{Global error after \(n\) Trotter steps}

We now pass from the one-step error to the total error after \(n\) steps. Writing
\begin{equation}
    V(\tau) = E(\Delta\tau)^n,
    \qquad
    V_{\mathrm{Trotter}}(\tau) = S(\Delta\tau)^n,
\end{equation}
we use the telescoping identity
\begin{equation}
    S^n - E^n
    =
    \sum_{m=0}^{n-1}
    S^{n-1-m}(S-E)E^m.
\end{equation}
Taking norms and applying the triangle inequality together with submultiplicativity gives
\begin{equation}
    \|S^n - E^n\|
    \le
    \sum_{m=0}^{n-1}
    \|S^{\,n-1-m}\|\,\|S-E\|\,\|E^m\|.
\end{equation}
Using
\begin{equation}
    \|S^{\,n-1-m}\| \le \|S\|^{n-1-m},
    \qquad
    \|E^m\| \le \|E\|^m,
\end{equation}
we find
\begin{equation}
    \|S^n - E^n\|
    \le
    \sum_{m=0}^{n-1}
    \|S\|^{n-1-m}\,\|E\|^m\,\|S-E\|.
\end{equation}
Bounding each term in the sum by the maximum of \(\|S\|\) and \(\|E\|\), we obtain
\begin{equation}
    \|S^n - E^n\|
    \le
    n\,\max(\|S\|,\|E\|)^{n-1}\,\|S-E\|.
\end{equation}

It remains to bound \(\|S\|\) and \(\|E\|\). For the exact step,
\begin{equation}
    \|E(\Delta\tau)\|
    =
    \left\|e^{-\frac{\Delta\tau}{2}H}\right\|
    \le
    e^{\frac{\Delta\tau}{2}\|H\|}.
\end{equation}
For the Trotter step,
\begin{align}
    \|S(\Delta\tau)\|
    &=
    \left\|
        \prod_{j=1}^{L} e^{A_j}
    \right\|
    \le
    \prod_{j=1}^{L} \|e^{A_j}\|
    \le
    \prod_{j=1}^{L} e^{\|A_j\|}
    \\
    &=
    \exp\left(\sum_{j=1}^{L}\|A_j\|\right)
    =
    \exp\left(
        \frac{\Delta\tau}{2}\sum_{j=1}^{L} |\alpha_j|\,\|h_j\|
    \right)
    =
    \exp\left(
        \frac{\Delta\tau}{2}\sum_{j=1}^{L} |\alpha_j|
    \right).
\end{align}
Define
\begin{equation}
    \mu := \sum_{j=1}^{L} |\alpha_j|.
\end{equation}
Since each \(h_j\) is a Pauli string and therefore satisfies \(\|h_j\|=1\), the triangle inequality implies
\begin{equation}
    \|H\|
    =
    \left\|\sum_{j=1}^{L} \alpha_j h_j\right\|
    \le
    \sum_{j=1}^{L} |\alpha_j|\,\|h_j\|
    =
    \mu.
\end{equation}
It follows that
\begin{equation}
    \|E(\Delta\tau)\| \le e^{\frac{\Delta\tau}{2}\mu},
    \qquad
    \|S(\Delta\tau)\| \le e^{\frac{\Delta\tau}{2}\mu}.
\end{equation}
Consequently,
\begin{equation}
    \max(\|S\|,\|E\|)^{n-1}
    \le
    \exp\left(
        \frac{(n-1)\Delta\tau}{2}\mu
    \right)
    \le
    e^{\frac{\tau}{2}\mu}.
\end{equation}
Combining this with the local-error estimate gives
\begin{equation}
    \|V_{\mathrm{Trotter}}(\tau)-V(\tau)\|
    \le
    n\,e^{\frac{\tau}{2}\mu}
    \left(
        \frac{\Delta\tau^2}{8}C
        +
        \mathcal{O}(\Delta\tau^3)
    \right).
\end{equation}
Since \(n=\tau/\Delta\tau\), we conclude
\begin{equation}
    \boxed{
    \|V_{\mathrm{Trotter}}(\tau)-V(\tau)\|
    \le
    e^{\frac{\tau}{2}\mu}
    \left(
        \frac{\tau\,\Delta\tau}{8}C
        +
        \mathcal{O}(\tau\,\Delta\tau^2)
    \right).
    }
\end{equation}
Equivalently, the first-order Trotter error obeys
\begin{equation}
    \boxed{
    \|V_{\mathrm{Trotter}}(\tau)-V(\tau)\|
    =
    \mathcal{O}\!\left(
        e^{\frac{\tau}{2}\mu}\, C\, \tau\, \Delta\tau
    \right).
    }
\end{equation}

\subsection{Discussion}

The above result shows that the first-order Trotter formula retains the expected global scaling obtained by accumulating a local error of order \(\mathcal{O}(\Delta\tau^2)\) over \(n=\tau/\Delta\tau\) steps. In particular, the global error scales as
\begin{equation}
    \|V_{\mathrm{Trotter}}(\tau)-V(\tau)\|
    =
    \mathcal{O}\!\left(
        e^{\frac{\tau}{2}\mu}\, C\, \tau\, \Delta\tau
    \right),
\end{equation}
where \(C = \sum_{j<k} |\alpha_j\alpha_k|\,\|[h_j,h_k]\|\) quantifies the non-commutativity of the Hamiltonian terms. In the special case where all Hamiltonian terms commute, one has \(C=0\), and the first-order Trotter formula becomes exact.

\textbf{Remark on scaling with fixed imaginary time.}
In many practical settings, the total imaginary time \(\tau\) is treated as fixed, while the Trotter step size \(\Delta\tau\) is reduced by increasing the number of steps \(n=\tau/\Delta\tau\). In this regime, the above bound implies linear scaling with the step size,
\begin{equation}
    \|V_{\mathrm{Trotter}}(\tau)-V(\tau)\|
    =
    \mathcal{O}(\Delta\tau),
\end{equation}
with a prefactor depending on \(\tau\), the Hamiltonian coefficients, and the commutator structure of the Hamiltonian.

\section{Error of the Normalized Trotterized Thermal State}
\label{app:trotter_imag_time_norm}

In this appendix, we derive a first-order error bound for the normalized imaginary-time evolved state associated with a Hamiltonian
\begin{equation}
    H = \sum_{j=1}^{m} \alpha_j h_j,
\end{equation}
where each \(h_j\) is a Pauli string and \(\alpha_j \in \mathbb{R}\). We consider the exact imaginary-time evolution operator
\begin{equation}
    V(\tau) = e^{-\frac{\tau}{2}H},
\end{equation}
and its first-order Trotter approximation
\begin{equation}
    V_{\mathrm{Trotter}}(\tau)
    =
    \left(
        \prod_{j=1}^{m}
        e^{-\frac{\Delta\tau}{2}\alpha_j h_j}
    \right)^n,
    \qquad
    \tau = n \Delta\tau.
\end{equation}

Given an initial density matrix \(\rho_0\), we define the exact and Trotterized normalized states
\begin{equation}
    \rho_{\mathrm{ex}}
    =
    \frac{V^\dagger \rho_0 V}
    {\Tr\!\left(V^\dagger \rho_0 V\right)},
    \qquad
    \rho_{\mathrm{tr}}
    =
    \frac{V_{\mathrm{Trotter}}^\dagger \rho_0 V_{\mathrm{Trotter}}}
    {\Tr\!\left(V_{\mathrm{Trotter}}^\dagger \rho_0 V_{\mathrm{Trotter}} \right)}.
\end{equation}

Our goal is to bound the error
\begin{equation}
    \bigl\|
        \rho_{\mathrm{ex}} - \rho_{\mathrm{tr}}
    \bigr\|_1,
\end{equation}
in terms of the Trotter step size \(\Delta\tau\).

Throughout, we use both the operator norm \(\|\cdot\|\) and the trace norm \(\|\cdot\|_1\), which satisfy
\begin{equation}
    \|XY\|_1 \le \|X\|\,\|Y\|_1,
    \qquad
    \|XYZ\|_1 \le \|X\|\,\|Y\|_1\,\|Z\|,
\end{equation}
together with
\begin{equation}
    \|X\| \le \|X\|_1.
\end{equation}
We also use the first-order Trotter bound established in Appendix~\ref{app:trotter_imag_time}. In its explicit form, this bound reads
\begin{equation}
    \|V(\tau) - V_{\mathrm{Trotter}}(\tau)\|
    \le
    e^{\frac{\tau}{2}\mu}\,
    \frac{\tau\,\Delta\tau}{8}
    \sum_{j<k} |\alpha_j\alpha_k|\,\|[h_j,h_k]\|,
\end{equation}
where
\begin{equation}
    \mu := \sum_{j=1}^m |\alpha_j|.
\end{equation}
Equivalently,
\begin{equation}
    \|V(\tau) - V_{\mathrm{Trotter}}(\tau)\|
    =
    \mathcal{O}\!\left(
        e^{\frac{\tau}{2}\mu}\, C\, \tau\, \Delta\tau
    \right),
\end{equation}
where
\begin{equation}
    C := \sum_{j<k} |\alpha_j\alpha_k|\,\|[h_j,h_k]\|.
\end{equation}

In particular, for fixed \(\tau\), sufficiently small \(\Delta\tau\), and fixed Hamiltonian data, the prefactor is independent of \(\Delta\tau\), so this implies the simpler bound
\begin{equation}
    \|V(\tau) - V_{\mathrm{Trotter}}(\tau)\|
    \le C_T\,\Delta\tau,
\end{equation}
for some constant \(C_T>0\).
\subsection{Unnormalized operators and positivity}

Define the unnormalized operators
\begin{equation}
    A := V^\dagger \rho_0 V,
    \qquad
    B := V_{\mathrm{Trotter}}^\dagger \rho_0 V_{\mathrm{Trotter}}.
\end{equation}
Then
\begin{equation}
    \rho_{\mathrm{ex}} = \frac{A}{\Tr A},
    \qquad
    \rho_{\mathrm{tr}} = \frac{B}{\Tr B}.
\end{equation}

Since \(\rho_0 \ge 0\), both \(A\) and \(B\) are positive semidefinite. Indeed, for any \(|\psi\rangle\),
\begin{equation}
    \langle \psi|A|\psi\rangle
    =
    \langle V\psi|\rho_0|V\psi\rangle
    \ge 0,
    \qquad
    \langle \psi|B|\psi\rangle
    =
    \langle V_{\mathrm{Trotter}}\psi|\rho_0|V_{\mathrm{Trotter}}\psi\rangle
    \ge 0.
\end{equation}
It follows that
\begin{equation}
    \|A\|_1 = \Tr A,
    \qquad
    \|B\|_1 = \Tr B.
\end{equation}

We first bound the difference \(A - B\):
\begin{equation}
    A - B
    =
    V^\dagger \rho_0 V
    -
    V_{\mathrm{Trotter}}^\dagger \rho_0 V_{\mathrm{Trotter}}.
\end{equation}
Adding and subtracting \(V_{\mathrm{Trotter}}^\dagger \rho_0 V\), we write
\begin{equation}
    A - B
    =
    (V^\dagger - V_{\mathrm{Trotter}}^\dagger)\rho_0 V
    +
    V_{\mathrm{Trotter}}^\dagger \rho_0 (V - V_{\mathrm{Trotter}}).
\end{equation}
Taking the trace norm and using the triangle inequality,
\begin{equation}
    \|A-B\|_1
    \le
    \left\|(V^\dagger - V_{\mathrm{Trotter}}^\dagger)\rho_0 V\right\|_1
    +
    \left\|V_{\mathrm{Trotter}}^\dagger \rho_0 (V - V_{\mathrm{Trotter}})\right\|_1.
\end{equation}
Using the bound
\begin{equation}
    \|XYZ\|_1 \le \|X\| \, \|Y\|_1 \, \|Z\|,
\end{equation}
we obtain
\begin{equation}
    \|A-B\|_1
    \le
    \|V^\dagger - V_{\mathrm{Trotter}}^\dagger\|\,\|\rho_0\|_1\,\|V\|
    +
    \|V_{\mathrm{Trotter}}^\dagger\|\,\|\rho_0\|_1\,\|V - V_{\mathrm{Trotter}}\|.
\end{equation}
Since
\begin{equation}
    \|X^\dagger\| = \|X\|,
    \qquad
    \|V^\dagger - V_{\mathrm{Trotter}}^\dagger\|
    =
    \|V - V_{\mathrm{Trotter}}\|,
\end{equation}
and \(\|\rho_0\|_1 = 1\), this simplifies to
\begin{equation}
    \|A-B\|_1
    \le
    \bigl(\|V\| + \|V_{\mathrm{Trotter}}\|\bigr)
    \|V - V_{\mathrm{Trotter}}\|.
\end{equation}

Hence there exists a constant \(M > 0\) such that
\begin{equation}
    \|V\| + \|V_{\mathrm{Trotter}}\| \le M.
\end{equation}
Therefore,
\begin{equation}
    \|A - B\|_1
    \le
    M\,\|V - V_{\mathrm{Trotter}}\|.
\end{equation}
Using the first-order Trotter bound from Appendix~\ref{app:trotter_imag_time}, we obtain
\begin{equation}
    \|A - B\|_1
    \le
    M\,e^{\frac{\tau}{2}\mu}\,
    \frac{\tau\,\Delta\tau}{8}
    \sum_{j<k} |\alpha_j\alpha_k|\,\|[h_j,h_k]\|.
\end{equation}
Equivalently,
\begin{equation}
    \|A - B\|_1
    =
    \mathcal{O}\!\left(
        e^{\frac{\tau}{2}\mu}\, C\, \tau\, \Delta\tau
    \right),
\end{equation}
where
\begin{equation}
    C := \sum_{j<k} |\alpha_j\alpha_k|\,\|[h_j,h_k]\|,
    \qquad
    \mu := \sum_{j=1}^m |\alpha_j|.
\end{equation}
This expression makes explicit how the error depends on the total imaginary time \(\tau\), the Hamiltonian coefficients, and the commutator structure.

In particular, if \(\tau\) is fixed and \(\Delta\tau \to 0\), then all prefactors are independent of \(\Delta\tau\), and the bound simplifies to
\begin{equation}
    \|A - B\|_1 = \mathcal{O}(\Delta\tau).
\end{equation}

\subsection{Normalization error}

Let
\begin{equation}
    a := \Tr A,
    \qquad
    b := \Tr B.
\end{equation}
Then
\begin{equation}
    \rho_{\mathrm{ex}} - \rho_{\mathrm{tr}}
    =
    \frac{A}{a} - \frac{B}{b}.
\end{equation}
Rewriting the difference gives
\begin{equation}
    \rho_{\mathrm{ex}} - \rho_{\mathrm{tr}}
    =
    \frac{A-B}{a}
    +
    \left(\frac{1}{a}-\frac{1}{b}\right)B.
\end{equation}
Taking the trace norm and using the triangle inequality, we obtain
\begin{equation}
    \|\rho_{\mathrm{ex}} - \rho_{\mathrm{tr}}\|_1
    \le
    \frac{\|A-B\|_1}{a}
    +
    \left|\frac{1}{a}-\frac{1}{b}\right|\,\|B\|_1.
\end{equation}
Since \(B \ge 0\), we have \(\|B\|_1 = \Tr B = b\). Therefore
\begin{equation}
    \|\rho_{\mathrm{ex}} - \rho_{\mathrm{tr}}\|_1
    \le
    \frac{\|A-B\|_1}{a}
    +
    \frac{|a-b|}{a}.
\end{equation}
Also,
\begin{equation}
    |a-b|
    =
    |\Tr(A-B)|
    \le
    \|A-B\|_1.
\end{equation}
Hence
\begin{equation}
    \|\rho_{\mathrm{ex}} - \rho_{\mathrm{tr}}\|_1
    \le
    \frac{2}{a}\,\|A-B\|_1.
\end{equation}

It remains to bound \(a\) from below. Since \(H\) is Hermitian, the exact imaginary-time evolution operator satisfies
\begin{equation}
    V = e^{-\frac{\tau}{2}H} = V^\dagger.
\end{equation}
Therefore
\begin{equation}
    A = V \rho_0 V,
\end{equation}
and by cyclicity of the trace,
\begin{equation}
    a
    =
    \Tr(V \rho_0 V)
    =
    \Tr(\rho_0 V^2)
    =
    \Tr(\rho_0 e^{-\tau H}).
\end{equation}
Let \(\lambda_{\max}(H)\) denote the largest eigenvalue of \(H\). Since
\begin{equation}
    e^{-\tau H} \ge e^{-\tau \lambda_{\max}(H)} I
\end{equation}
as operators, it follows that
\begin{equation}
    a
    =
    \Tr(\rho_0 e^{-\tau H})
    \ge
    e^{-\tau \lambda_{\max}(H)} \Tr(\rho_0)
    =
    e^{-\tau \lambda_{\max}(H)}.
\end{equation}
Thus \(a\) is bounded below by a positive constant independent of \(\Delta\tau\). Combining this with the previous estimate gives
\begin{equation}
    \|\rho_{\mathrm{ex}} - \rho_{\mathrm{tr}}\|_1
    \le
    2 e^{\tau \lambda_{\max}(H)} \|A-B\|_1.
\end{equation}
Using the bound on the unnormalized error, we obtain
\begin{equation}
    \|\rho_{\mathrm{ex}} - \rho_{\mathrm{tr}}\|_1
    =
    \mathcal{O}\!\left(
        e^{\tau \lambda_{\max}(H)}
        e^{\frac{\tau}{2}\mu}
        C\,\tau\,\Delta\tau
    \right),
\end{equation}
where
\begin{equation}
    \mu := \sum_{j=1}^m |\alpha_j|,
    \qquad
    C := \sum_{j<k} |\alpha_j\alpha_k|\,\|[h_j,h_k]\|.
\end{equation}
In particular, for fixed \(\tau\),
\begin{equation}
    \|\rho_{\mathrm{ex}} - \rho_{\mathrm{tr}}\|_1
    =
    \mathcal{O}(\Delta\tau).
\end{equation}

\subsection{Final error bound and discussion}

Combining the normalization estimate
\begin{equation}
    \|\rho_{\mathrm{ex}} - \rho_{\mathrm{tr}}\|_1
    \le
    \frac{2}{a}\,\|A-B\|_1,
\end{equation}
with the lower bound
\begin{equation}
    a \ge e^{-\tau \lambda_{\max}(H)},
\end{equation}
we obtain
\begin{equation}
    \|\rho_{\mathrm{ex}} - \rho_{\mathrm{tr}}\|_1
    \le
    2 e^{\tau \lambda_{\max}(H)} \|A-B\|_1.
\end{equation}
Using the bound on the unnormalized error, this yields
\begin{equation}
    \|\rho_{\mathrm{ex}} - \rho_{\mathrm{tr}}\|_1
    =
    \mathcal{O}\!\left(
        e^{\tau \lambda_{\max}(H)}
        e^{\frac{\tau}{2}\mu}
        C\,\tau\,\Delta\tau
    \right),
\end{equation}
where
\begin{equation}
    \mu := \sum_{j=1}^m |\alpha_j|,
    \qquad
    C := \sum_{j<k} |\alpha_j\alpha_k|\,\|[h_j,h_k]\|.
\end{equation}

This bound makes explicit the dependence of the error on the total imaginary time \(\tau\), the Hamiltonian coefficients, and the non-commutativity of the Hamiltonian terms. In particular, for fixed \(\tau\), all prefactors are independent of \(\Delta\tau\), and the error scales linearly with the Trotter step size:
\begin{equation}
    \boxed{
    \|\rho_{\mathrm{ex}} - \rho_{\mathrm{tr}}\|_1
    =
    \mathcal{O}(\Delta\tau).
    }
\end{equation}
Since \(\|X\| \le \|X\|_1\), the same scaling holds for the operator norm.

\textbf{Remark (thermal-state construction).}
For the choice \(\rho_0 = \mathbb{I}/2^N\), one obtains
\begin{equation}
    \rho_{\mathrm{ex}}
    =
    \frac{e^{-\tau H}}{\Tr(e^{-\tau H})},
    \qquad
    \rho_{\mathrm{tr}}
    =
    \frac{
        V_{\mathrm{Trotter}}^\dagger \left(\frac{\mathbb{I}}{2^N}\right) V_{\mathrm{Trotter}}
    }{
        \Tr\!\left(
            V_{\mathrm{Trotter}}^\dagger \left(\frac{\mathbb{I}}{2^N}\right) V_{\mathrm{Trotter}}
        \right)
    }.
\end{equation}
In this case, the above bound implies
\begin{equation}
    \left\|
        \frac{e^{-\tau H}}{\Tr(e^{-\tau H})}
        -
        \frac{
            V_{\mathrm{Trotter}}^\dagger \left(\frac{\mathbb{I}}{2^N}\right) V_{\mathrm{Trotter}}
        }{
            \Tr\!\left(
                V_{\mathrm{Trotter}}^\dagger \left(\frac{\mathbb{I}}{2^N}\right) V_{\mathrm{Trotter}}
            \right)
        }
    \right\|_1
    =
    \mathcal{O}\!\left(
        e^{\tau \lambda_{\max}(H)}
        e^{\frac{\tau}{2}\mu}
        C\,\tau\,\Delta\tau
    \right),
\end{equation}
where
\begin{equation}
    \mu := \sum_{j=1}^m |\alpha_j|,
    \qquad
    C := \sum_{j<k} |\alpha_j\alpha_k|\,\|[h_j,h_k]\|.
\end{equation}
In particular, for fixed \(\tau\), this reduces to
\begin{equation}
    \left\|
        \frac{e^{-\tau H}}{\Tr(e^{-\tau H})}
        -
        \frac{
            V_{\mathrm{Trotter}}^\dagger \left(\frac{\mathbb{I}}{2^N}\right) V_{\mathrm{Trotter}}
        }{
            \Tr\!\left(
                V_{\mathrm{Trotter}}^\dagger \left(\frac{\mathbb{I}}{2^N}\right) V_{\mathrm{Trotter}}
            \right)
        }
    \right\|_1
    =
    \mathcal{O}(\Delta\tau).
\end{equation}
\end{document}